\documentclass[aps,prb,reprint,superscriptaddress, preprintnumbers]{revtex4-2}

\usepackage{graphicx}
\usepackage{blindtext}
\usepackage{amsmath,amssymb}
\usepackage{color}
\usepackage{bm}
\usepackage{bbm}
\usepackage{mathtools, physics}
\usepackage{amsthm}
\usepackage[normalem]{ulem}
\usepackage[colorlinks=true,citecolor=blue,linkcolor=blue]{hyperref}

\newcommand\Img[2][]{\vcenter{\hbox{\includegraphics[scale=#1]{#2}}}}

\newcommand{\lp}{\lparen}
\newcommand{\rp}{\rparen}
\newcommand{\lvec}[1]{\mathrlap{\reflectbox{\ensuremath{\vec{\phantom{#1}}}}}#1}

\begin{document}
\title{$c=-2$ conformal field theory in quadratic band touching}
\author{Rintaro Masaoka}
\affiliation{Department of Applied Physics, The University of Tokyo, Tokyo 113-8656, Japan}
\begin{abstract}
Quadratic band touching in fermionic systems defines a universality class distinct from that of linear Dirac points, yet its characterization as a quantum critical point remains incomplete.
In this work, I show that a $(d+1)$-dimensional free-fermion model with quadratic band touching exhibits spatial conformal invariance, and that its equal-time ground-state correlation functions are exactly captured by the $d$-dimensional symplectic fermion theory.
I establish this correspondence by constructing explicit mappings between physical fermionic operators and the fields of the symplectic fermion theory.
I further explore the implications of this correspondence in two spatial dimensions, where the symplectic fermion theory is a logarithmic conformal field theory with central charge $c=-2$.
In the corresponding $(2+1)$-dimensional systems, I identify anyonic excitations originating from the underlying symplectic fermion theory, even though the Hamiltonian is gapless.
Transporting these excitations along non-contractible loops generates transitions among topologically degenerate ground states, in close analogy with those in topologically ordered phases.
Moreover, the action of a $2\pi$ rotation on these excitations is represented by a Jordan block, reflecting the logarithmic character of the associated conformal field theory.
\end{abstract}

\maketitle

\section{Introduction}
\label{sec: introduction}
Quadratic band touching (QBT) with non-relativistic dispersion $\varepsilon\propto\pm|\bm{k}|^2$ in fermion systems defines a distinct low-energy universality class from linear Dirac points with relativistic dispersion $\varepsilon\propto\pm|\bm{k}|$. 

In particular, in two spatial dimensions QBT has attracted considerable attention as a platform for studying interaction-driven phases. In non-interacting systems, a QBT point is protected by time-reversal symmetry together with fourfold or sixfold rotational symmetry $C_4$ or $C_6$~\cite{sunTimereversalSymmetryBreaking2008}.
In interacting systems, renormalization-group analyses have shown that, unlike Dirac points, QBT points are marginally unstable against weak short-range interactions~\cite{sunTopologicalInsulatorsNematic2009,uebelackerInstabilitiesQuadraticBand2011,murrayRenormalizationGroupStudy2014}.
This instability makes QBT systems fertile ground for exploring interaction-driven phases.
Theoretical studies have proposed nematic order, quantum anomalous Hall, and quantum spin Hall states emerging from weak interactions near QBT.
QBT with interaction effects has been studied in various platforms, including Bernal-stacked bilayer graphene~\cite{mccannLandauLevelDegeneracy2006,koshinoTransportBilayerGraphene2006,pujariInteractionInducedDiracFermions2016}, checkerboard lattice~\cite{sunTimereversalSymmetryBreaking2008,sunTopologicalInsulatorsNematic2009,zengTuningTopologicalPhase2018}, kagome lattice~\cite{liuSpontaneousSymmetryBreaking2010,zhuInteractionDrivenSpontaneousQuantum2016}, and optical lattice realizations~\cite{olschlagerTopologicallyInducedAvoided2012,sunTopologicalSemimetalFermionic2012,liPhysicsHigherOrbital2016}.
However, before considering interactions, it is essential to fully understand the properties of non-interacting QBT systems. 

In this work I revisit non-interacting QBT as a quantum critical point. Although such free-fermion models are straightforward to solve, an essential aspect has been overlooked: their equal-time ground-state correlation functions exhibit spatial conformal invariance governed by the symplectic fermion theory.

Conformal field theory (CFT)~\cite{difrancescoConformalFieldTheory1997,ginspargAppliedConformalField1988} provides a powerful framework for describing critical phenomena.
However, it cannot be directly applied to non-relativistic systems exhibiting anisotropic scaling $t \mapsto \lambda^z t, x \mapsto \lambda x$ with $z \ne 1$, because it intrinsically assumes Lorentz invariance.
Nevertheless, certain non-relativistic quantum critical points, known as conformal quantum critical points (CQCPs)~\cite{ardonneTopologicalOrderConformal2004}, exhibit spatial conformal invariance in their ground-state correlation functions, even though they do not possess full space-time conformal invariance.

The concept of CQCP originally emerged in the context of quantum dimer models and so-called Rokhsar-Kivelson (RK) states~\cite{rokhsarSuperconductivityQuantumHardCore1988}.
In a generalized formulation~\cite{henleyClassicalQuantumDynamics2004,castelnovoQuantumMechanicsClassical2005}, RK states are constructed as
\begin{align}
| \mathrm{RK} \rangle = \frac1{\sqrt{Z}} \sum_{\mathcal{C}} \sqrt{e^{-\beta E(\mathcal{C})}} | \mathcal{C} \rangle,
\end{align}
where $| \mathcal{C} \rangle$ represents orthonormal basis states labeled by classical configurations $\mathcal{C}$ of a classical statistical model with energy $E(\mathcal{C})$ at inverse temperature $\beta$.
The normalization constant $Z = \sum_{\mathcal{C}} e^{-\beta E(\mathcal{C})}$ can be interpreted as the partition function of the classical statistical model.
RK states form a subclass of tensor-network states, and explicit representations as projected entangled pair states are known~\cite{verstraeteCriticalityAreaLaw2006}.

An important property of RK states is the quantum-classical correspondence, namely that their equal-time ground-state correlation functions coincide with correlation functions of the corresponding classical statistical model.
For the operator $\hat F$ that is diagonal in the basis $| \mathcal{C} \rangle$ as $\hat F| \mathcal{C} \rangle = F(\mathcal{C})| \mathcal{C} \rangle$, the correspondence is expressed as
\begin{align}
\langle \mathrm{RK} | \hat F | \mathrm{RK} \rangle
= \frac1{Z} \sum_{\mathcal{C}} F(\mathcal{C})e^{-\beta E(\mathcal{C})} = \langle F(\mathcal{C}) \rangle.
\label{RK correspondence}
\end{align}
The parent Hamiltonians for such RK states, called generalized RK Hamiltonians, realize CQCPs when the corresponding classical statistical model is at a critical point~\cite{ardonneTopologicalOrderConformal2004, verstraeteCriticalityAreaLaw2006, isakovDynamicsConformalQuantum2011}.
From a field-theoretic viewpoint, such a construction of dynamics from classical statistical models can be formulated as stochastic quantization~\cite{parisiPERTURBATIONTHEORYGAUGE1981,dijkgraafRelatingFieldTheories2010}.
While the framework of RK Hamiltonians is powerful for constructing CQCPs, it is limited to bosonic systems, because RK states defined through positive Boltzmann weights cannot accommodate negative signs arising from fermionic statistics
~\footnote{Although some statistical models, such as the Ising and dimer models, admit fermionic representations, the corresponding RK Hamiltonians are constructed on a bosonized basis of classical configurations.}.

Let us now return to QBT systems.
In this work, I show that the ground states of $(d+1)$-dimensional QBT systems exhibit a correspondence, analogous to Eq.~\eqref{RK correspondence}, with the $d$-dimensional symplectic fermion theory~\cite{kauschCuriositiesC21995, kauschSymplecticFermions2000}.
Consequently, QBT systems realize a CQCP, even though their ground states cannot be written as RK states.
The symplectic fermion theory is defined in terms of two Grassmann fields $\theta$ and $\theta^*$ with the action
\begin{align}
S[\theta, \theta^*] = \frac1{4\pi}\int\mathrm{d}^dx \partial_i\theta(\bm{x})\partial^i\theta^*(\bm{x}).
\label{action of SF (intro)}
\end{align}
The explicit correspondence between the physical fermionic operators and the symplectic fermion fields is given by
\begin{align}
\hat{\psi}_i \leftrightarrow \frac{\partial_i\theta}{\sqrt{4\pi}}, \quad \hat{\psi}^\dagger_i \leftrightarrow \frac{\partial_i\theta^*}{\sqrt{4\pi}}.
\end{align}

QBT systems have been regarded as a fermionic analogue of the quantum Lifshitz model, which is a prototypical CQCP described by a free scalar field~\cite{fradkinFieldTheoriesCondensed2013}.
One reason is that they are free theories sharing the same dynamical exponent $z=2$.
Furthermore, the relation between QBT and the quantum Lifshitz model is parallel to that between the Dirac fermion and the Klein-Gordon theory~\cite{fradkinFieldTheoriesCondensed2013}.
Beyond these insights, my result provides a precise understanding of similarities and differences between these two models.
While both models are CQCPs, their underlying CFTs differ: the quantum Lifshitz model corresponds to a free boson CFT, whereas QBT systems correspond to the symplectic fermion CFT.

In the latter part of this work, I explore the consequences of the established correspondence in two spatial dimensions.
The two-dimensional symplectic fermion theory exhibits several intriguing properties.
In particular, it is a well-known example of a logarithmic CFT~\cite{gurarieLogarithmicOperatorsConformal1993,flohrBitsPiecesLogarithmic2003,creutzigLogarithmicConformalField2013}, where logarithmic terms appear in correlation functions due to Jordan block structures in representations of the Virasoro algebra.
A natural question is whether such behavior can also be observed in $(2+1)$-dimensional QBT systems.
I show that $(2+1)$-dimensional QBT systems indeed host anyonic excitations originating from the underlying symplectic fermion, even though the Hamiltonians are gapless.
I explicitly construct three types of localized excitations, which I call Dirichlet, Neumann, and composite excitations.
Moving these excitations along non-contractible loops implements transitions between topologically degenerate ground states, in close analogy with topologically ordered phases~\cite{kitaevFaulttolerantQuantumComputation2003,levinStringnetCondensationPhysical2005,kitaevAnyonsExactlySolved2006,nayakNonAbelianAnyonsTopological2008,simonTopologicalQuantum2023}.
I further analyze the spins of these anyons and demonstrate that they exhibit a Jordan block structure, which can be understood from the logarithmic nature of the underlying symplectic fermion theory.

The remainder of this paper is organized as follows.
In Sec.~\ref{sec: correspondence with symplectic fermion}, I introduce a continuum $(d+1)$-dimensional free-fermion model with QBT and show that its ground-state correlation functions correspond to those of the $d$-dimensional symplectic fermion theory.
In Sec.~\ref{sec: lattice model}, I discuss the same correspondence in lattice models and provide an explicit example on the square lattice.
From Sec.~\ref{sec: review of symplectic fermion} onward, I restrict the discussion to two spatial dimensions.
In Sec.~\ref{sec: review of symplectic fermion}, we briefly review symplectic fermion theory as a logarithmic CFT.
In Sec.~\ref{sec: anyons}, I first construct anyonic excitations.
I use them to explain the topological degeneracy of the QBT model, and examine the spins of these anyons, revealing their Jordan block structure.
Finally, in Sec.~\ref{sec: discussion and outlook}, I conclude with a summary and an outlook.

\section{Correspondence with symplectic fermion theory}
\label{sec: correspondence with symplectic fermion}

In this section, I introduce $(d+1)$-dimensional free-fermion continuum models with QBT and demonstrate that their ground-state correlation functions correspond to those of the symplectic fermion theory.

\subsection{Definition of the continuum model}
I consider a $(d+1)$-dimensional continuum model of $d$-component fermions with QBT.
While these components might represent spin or sublattice degrees of freedom, I express them as 1-form fermions for clarity of notation.
The annihilation and creation operators are expressed as
\begin{align}
\hat{\psi}(\bm{x}) = \hat{\psi}_i(\bm{x})dx^i,\quad
\hat{\psi}^\dagger(\bm{x}) = \hat{\psi}^\dagger_i(\bm{x})dx^i,
\end{align}
where $i = 1, 2, ..., d$ and $\bm{x} = (x^1, x^2, ..., x^d)$.
These satisfy the canonical anticommutation relations
\begin{align}
\{\hat{\psi}_i(\bm{x}), \hat{\psi}^\dagger_j(\bm{y})\} = \delta_{ij}\delta(\bm{x}-\bm{y}).
\end{align}
The Hamiltonian of the continuum model is given by
\begin{align}
\hat{H}
&
= t_+\lp d\hat{\psi}^\dagger, d\hat{\psi} \rp + t_-\lp \delta\hat{\psi}, \delta\hat{\psi}^\dagger \rp \nonumber\\
&
= \int(t_+d\hat{\psi}^\dagger(\bm{x}) \wedge\star d\psi(\bm{x}) + t_-\delta\hat{\psi}(\bm{x})\wedge\star\delta\hat{\psi}^\dagger(\bm{x})),
\label{def of Hamiltonian}
\end{align}
where $t_\pm$ are positive constants, $d$ is the exterior derivative, $\delta = -{\star}d{\star}$ is the codifferential, and $\star$ is the Hodge star operator.
The inner product of differential forms is defined as $(a, b) = \int a \wedge \star b$.

In this paper, I mainly focus on the case of $d=2$.
Then, Eq.~\eqref{def of Hamiltonian} gives a general effective Hamiltonian of a rotationally symmetric two-band system exhibiting QBT, up to unitary transformations of the internal degrees of freedom.
In two dimensions, the explicit forms of $d\hat{\psi}$ and $\delta\hat{\psi}$ are given by
\begin{gather}
d\hat{\psi}(\bm{x}) = (\partial_1\hat{\psi}_2(\bm{x}) - \partial_2\hat{\psi}_1(\bm{x}))dx^1\wedge dx^2, \\
\delta\hat{\psi}(\bm{x}) = - \partial_1\hat{\psi}_1(\bm{x}) - \partial_2\hat{\psi}_2(\bm{x}),
\end{gather}
and the same applies for $\hat{\psi}^\dagger$.
In momentum space, the Hamiltonian is expressed as
\begin{align}
\hat{H}
&
= \int\frac{\mathrm{d}^2\bm{k}}{(2\pi)^2}
\begin{pmatrix}
\hat{\psi}^\dagger_{1,\bm{k}} & \hat{\psi}^\dagger_{2,\bm{k}}
\end{pmatrix}
H(\bm{k})
\begin{pmatrix}
\hat{\psi}_{1,\bm{k}}  \\\hat{\psi}_{2,\bm{k}}
\end{pmatrix},
\\
H(\bm{k}) &
= \begin{pmatrix}
t_+k_2^2- t_-k_1^2 & (t_++t_-)k_1k_2 \\
(t_++t_-)k_2k_1 & t_+k_1^2- t_-k_2^2
\end{pmatrix}\nonumber \\
&
= \frac{t_+-t_-}{2}(k_1^2+k_2^2)\sigma_0
- \frac{t_++t_-}{2}(k_1^2-k_2^2)\sigma_z \nonumber\\
&\quad
 + (t_++t_-)k_1k_2\sigma_x.
\end{align}
Here, $\sigma_0$ is the $2\times2$ identity matrix and $\sigma_x, \sigma_z$ are Pauli matrices.
Introducing complex coordinates $k = k_1 + ik_2$ and $\bar{k} = k_1 - ik_2$, the Hamiltonian is rewritten as
\begin{align}
H(\bm{k}) = \frac{t'}{2}k\bar{k}\sigma_0 - \frac{t}{2}k^2\frac{\sigma_z+i\sigma_x}{2} - \frac{t}{2}\bar{k}^2\frac{\sigma_z-i\sigma_x}{2},
\end{align}
where $t \coloneqq t_++t_-$ and $t' \coloneqq t_+-t_-$.
By diagonalizing this Hamiltonian, the energy dispersions $\epsilon_\pm$ and the Bloch states $b_\pm$ are obtained as
\begin{align} &
\epsilon_+(\bm{k}) = t_+|\bm{k}|^2,\quad \vec{b}_+(\bm{k}) = \frac{\bm{k}^\bot}{|\bm{k}|}, \label{upper band} \\ &
\epsilon_-(\bm{k}) = -t_-|\bm{k}|^2,\quad \vec{b}_-(\bm{k}) = \frac{\bm{k}}{|\bm{k}|}, \label{lower band}
\end{align}
where $\bm{k}^\bot = (-k_2, k_1)$.
Thus, the band dispersion exhibits quadratic touching at $\bm{k} = \bm{0}$.
According to the quantum geometric characterization of band touchings~\cite{rhimClassificationFlatBands2019,rhimSingularFlatBands2021}, this band touching is referred to as singular because the Bloch states cannot be chosen continuously at the touching point.

Eqs.~\eqref{upper band} and \eqref{lower band} also hold for general spatial dimensions $d$.
In general dimensions, the upper bands have $(d-1)$-fold degeneracy corresponding to transverse modes, while the lower band is a non-degenerate longitudinal mode.

We have assumed that $t_\pm$ are positive, ensuring that the two bands have opposite sign dispersions.
If $t_\pm$ are both negative, we swap $\hat{\psi}$ and $\star\hat{\psi}^\dagger$ to recover the positive case.
If either $t_+$ or $t_-$ is zero, the Hamiltonian has a flat band at zero energy, and the subsequent analysis still holds if we assume half filling.
We exclude cases where $t_\pm$ have opposite signs, as the ground states become either almost fully occupied or empty in such cases.



\subsection{Ground state and path integral representation}

The ground state with all negative-energy modes occupied is expressed as
\begin{align}
| \mathrm{GS} \rangle
&
= \prod_{\bm{k}\ne\bm{0}} \frac{ik^j}{|\bm{k}|} \hat{\psi}_{j,\bm{k}}^\dagger| 0 \rangle \nonumber \\
&
=  \frac1{\sqrt{Z}} \prod_{\bm{k}\ne\bm{0}} \frac{ik^j}{\sqrt{4\pi}}\hat{\psi}_{j,\bm{k}}^\dagger| 0 \rangle,
\label{GS in k space}
\end{align}
where the index $j$ is summed over and $Z$ is a factor included for later convenience, defined as $Z = \prod_{\bm{k}\ne\bm{0}}(|\bm{k}|^2/4\pi)$.
To improve the clarity of equations, I define a shorthand notation $g \coloneqq 1/\sqrt{4\pi}$.

Let us represent the ground state in Eq.~\eqref{GS in k space} using a fermionic path integral.
For each non-zero mode, I insert a Grassmann variable $\theta_{\bm{k}}$ by the identity
\begin{align}
 x = \int\exp(x\theta_{\bm{k}})\lvec{\mathrm{d}}\theta_{\bm{k}}.
\end{align}
Here, I use the right derivative $\lvec{\mathrm{d}}\theta_{\bm{k}} \coloneqq \lvec{\partial}/\partial\theta_{\bm{k}}$ to avoid later sign complications.
Using this identity, the ground state is expressed as
\begin{align}
| \mathrm{GS} \rangle
&
= \frac1{\sqrt{Z}}\prod_{\bm{k}\ne\bm{0}}\left[\int\exp\left(\frac{ik^j}{\sqrt{4\pi}}\hat{\psi}_{j,\bm{k}}^\dagger\theta_{\bm{k}}\right)\lvec{\mathrm{d}}\theta_{\bm{k}}\right]| 0 \rangle \nonumber\\ &
= \frac1{\sqrt{Z}}\int\theta_{\bm{k}=\bm{0}}
\exp\left(-g\int\frac{\mathrm{d}^2\bm{k}}{(2\pi)^2}ik^j\theta_{\bm{k}}\hat{\psi}^\dagger_{j,\bm{k}}\right)| 0 \rangle\lvec{\mathcal{D}}\theta \nonumber\\ &
= \frac1{\sqrt{Z}}\int\theta_{\bm{k}=\bm{0}}\exp\left(-g\int\mathrm{d}^2\bm{x}\,\partial^j\theta(\bm{x})\hat{\psi}^\dagger_{j}(\bm{x})\right)| 0 \rangle\lvec{\mathcal{D}}\theta.
\end{align}
Thus, the ground state can be represented as
\begin{align}
| \xi \rangle \coloneqq \frac1{\sqrt{Z}}\int\xi| gd\theta \rangle_\mathrm{fc}\lvec{\mathcal{D}}\theta,
\label{path integral representation of GS}
\end{align}
where $\xi \coloneqq \theta_{\bm{k}=\bm{0}}$ is the zero mode and $| gd\theta \rangle_\mathrm{fc}$ is a fermionic coherent state given by
\begin{align}
 | gd\theta \rangle_\mathrm{fc}
 \coloneqq \exp\left( -g\int\mathrm{d}^d\bm{x}\, \partial^j\theta(\bm{x})\hat\psi^\dagger_{j}(\bm{x}) \right)| 0 \rangle.
\end{align}
More generally, I introduce a notation
\begin{align}
| X \rangle \coloneqq \frac1{\sqrt{Z}}\int X| gd\theta \rangle_\mathrm{fc}\lvec{\mathcal{D}}\theta,
\end{align}
for an arbitrary functional $X$ of $\theta$.

Other degenerate ground states can be constructed by acting the zero-mode creation operators $\hat{\psi}^\dagger_{i,\bm{k}=0}$ on $| \xi \rangle$.


\subsection{Quantum-classical correspondence}
The bra of the ground state in Eq.~\eqref{path integral representation of GS} is given by
\begin{align}
\langle \xi^* |
= \frac{1}{\sqrt{Z}}\int \mathcal{D}\theta^*\langle gd\theta^* |_\mathrm{fc}\xi^*,
\label{bra GS}
\end{align}
where $\xi^* = \theta^*_{\bm{k}=\bm{0}}$ and $\langle gd\theta^* |_\mathrm{fc}$ is a fermionic coherent state given by $\langle gd\theta^* |_\mathrm{fc} = | gd\theta^* \rangle_\mathrm{fc}^\dagger$.
Here, $\theta^*$ are not the complex conjugates of $\theta$, but independent Grassmann variables.
Then, the norm of the ground state is
\begin{align}
\langle \xi^* | \xi \rangle
&
= \frac1{Z}\int\mathcal{D}\theta^*\langle gd\theta^* |_\mathrm{fc}\xi^* \int\xi| gd\theta \rangle_\mathrm{fc}\lvec{\mathcal{D}}\theta \nonumber\\ &
= \frac1{Z}\int\mathcal{D}\theta^*\xi^*\xi\exp\left( g^2 \lp d\theta^*,d\theta \rp \right)\lvec{\mathcal{D}}\theta \nonumber\\ &
= \frac1{Z}\int\mathcal{D}\theta\mathcal{D}\theta^*\xi^*\xi\exp(-S[\theta,\theta^*]).
\end{align}
The normalization constant $Z$ can be regarded as a partition function, similar to the case of RK states.
The action $S[\theta, \theta^*]$ is given by
\begin{align}
S[\theta, \theta^*] = -g^2\lp d\theta^*, d\theta \rp = \frac1{4\pi} \int d^dx \partial_i\theta(\bm{x})\partial^i\theta^*(\bm{x}),
\end{align}
which coincides with that of the symplectic fermion theory.
This theory is conformally invariant for any $d\ge2$.
In two dimensions, this theory is known as a logarithmic conformal field theory with central charge $c=-2$~\cite{kauschCuriositiesC21995,kauschSymplecticFermions2000}.
Three-dimensional symplectic fermion theory has also received attention in the context of dS/CFT correspondence~\cite{ngStateOperatorCorrespondence2013,anninosHigherSpinRealization2016}.

The correlation functions in the QBT systems correspond exactly to those of the symplectic fermion. For the two-point function, we have
\begin{align}&
\langle \xi^* | \hat{\psi}^\dagger_i(\bm{x})\hat{\psi}_j(\bm{y}) | \xi \rangle
\nonumber\\
&
= \frac1{Z} \int\mathcal{D}\theta^*\xi^*\langle gd\theta^*|_\mathrm{fc} \hat{\psi}^\dagger_i(\bm{x})\hat{\psi}_j(\bm{y}) \int\xi | gd\theta \rangle_\mathrm{fc}\lvec{\mathcal{D}}\theta \nonumber\\ &
= \frac{g^2}{Z}\int\mathcal{D}\theta\mathcal{D}\theta^*\xi^* \partial_i\theta^*(\bm{x})\partial_j\theta(\bm{y})\xi e^{-S[\theta, \theta^*]}\nonumber\\ &
= \frac1{4\pi}\langle \xi^*\partial_i\theta^*(\bm{x})\partial_j\theta(\bm{y})\xi \rangle,
\end{align}
where we have defined
\begin{align}
\langle X \rangle \coloneqq \frac1{Z}\int\mathcal{D}\theta\mathcal{D}\theta^* Xe^{-S[\theta, \theta^*]}.
\end{align}
Notably, such expectation values satisfy $\langle 1 \rangle = 0$ due to the presence of zero modes in the Grassmann integration.
This behavior is a characteristic feature of logarithmic CFT.
Similarly, for general correlation functions, we have
\begin{align}
\langle \xi^* | F[\hat{\psi}^\dagger]G[\hat{\psi}] | \xi \rangle
&
= \langle \xi^*F[gd\theta^*]G[gd\theta]\xi \rangle,
\end{align}
for arbitrary functionals $F$ and $G$.
Thus, the correspondence between the QBT systems and the symplectic fermion theory is summarized as
\begin{align}
\hat{\psi} \leftrightarrow \frac{d\theta}{\sqrt{4\pi}},\quad
\hat{\psi}^\dagger \leftrightarrow \frac{d\theta^*}{\sqrt{4\pi}}.
\label{correspondence for QBT}
\end{align}
Note that, in addition to simply making this replacement in correlation functions, we need to insert the zero modes $\xi^*\xi$.

Multi-point correlation functions can be explicitly calculated by Gaussian integration as
\begin{align} 
&
\langle \xi^*|\hat{\psi}_{i_1}^\dagger(\bm{x}_1)\cdots\hat{\psi}_{i_n}^\dagger(\bm{x}_n)\hat{\psi}_{j_1}(\bm{y}_1)\cdots\hat{\psi}_{j_n}(\bm{y}_n) | \xi \rangle  \nonumber\\ &
= \frac1{(4\pi)^n}\langle \xi^*\partial_{i_1}\theta^*(\bm{x}_1)\cdots\partial_{i_n}\theta^*(\bm{x}_n)\partial_{j_1}\theta(\bm{y}_1)\cdots\partial_{j_n}\theta(\bm{y}_n)\xi \rangle \nonumber\\
&
= \det\left[\frac{\partial}{\partial x_{i_l}}\frac{\partial}{\partial y_{j_m}}\ln |\bm{x}-\bm{y}|^2 \right]_{1\le l,m\le n}.
\end{align}

\subsection{Frustration-freeness}
A key property of the QBT model is its frustration-freeness, which means that its ground states minimize each local term of the Hamiltonian simultaneously.
I will use this property in the later discussions on anyonic excitations.
I also note that frustration-freeness is crucial for ensuring the non-relativistic nature of the
dynamics in gapless systems~\cite{gossetLocalGapThreshold2016,gossetCorrelationLengthGap2016,masaokaRigorousLowerBound2024}.

For the QBT model, the frustration-free conditions are expressed as
\begin{align}&
d\hat{\psi}^\dagger(\bm{x})\wedge\star d\hat{\psi}(\bm{x})| \mathrm{GS} \rangle = 0,\\ &
\delta\hat{\psi}(\bm{x})\wedge\star\delta\hat{\psi}^\dagger(\bm{x})| \mathrm{GS} \rangle = 0.
\end{align}
Since the local terms have the positive semidefinite form $\hat A^\dagger\hat A$, these conditions are equivalent to
\begin{align}
d\hat{\psi}(\bm{x})| \mathrm{GS} \rangle = 0,\quad
\delta\hat{\psi}^\dagger(\bm{x})| \mathrm{GS} \rangle = 0.
\label{frustration-free condition}
\end{align}
In momentum space, these conditions are written as
\begin{align}
 (k_i\hat{\psi}_j - k_j\hat{\psi}_i)| \mathrm{GS} \rangle = 0,\quad
 k^i\hat{\psi}^\dagger_i| \mathrm{GS} \rangle = 0,
\end{align}
which can be easily verified from the Bloch states in Eqs.~\eqref{upper band} and \eqref{lower band}.

I note that writing down a condition similar to Eq.~\eqref{frustration-free condition} is possible for general free-fermion systems~\cite{sengokuQuasilocalFrustrationFreeFree2025}.
The essential implication of frustration-freeness is the locality of operators $d$ and $\delta$, which in turn ensure the locality of the corresponding Euclidean field theory.

\section{Lattice realizations}
\label{sec: lattice model}

In this section, I construct lattice free-fermion models with QBT that show the same correspondence with the symplectic fermion theory as in the continuum case.
The key step is to replace differential forms by their counterparts in the discrete language of homology theory.

\subsection{Notation from homology theory}

We use the language of homology theory to describe lattice models with QBT.
We work with $d$-dimensional lattices with vertices, edges, and faces.
Higher-dimensional cells will not be needed in the following discussion.
Let $V = \{v\}$ denote the set of vertices, $E = \{e\}$ the set of edges, and $F = \{f\}$ the set of faces.
Formal vector spaces $C_0, C_1, C_2$ spanned by these cells are defined as
\begin{align}&
C_0 = \operatorname{Span}\{v \mid v \in V\},\nonumber\\ &
C_1 = \operatorname{Span}\{e \mid e \in E\},\nonumber\\ &
C_2 = \operatorname{Span}\{f \mid f \in F\},
\end{align}
with suitable coefficients.
Each cell has an orientation, and reversing the orientation gives a negative sign.
Elements of $C_p$ are called $p$-chains.
The $p$-cochain space $C^p$ is defined as the dual space of $C_p$.
We identify $C^p$ with $C_p$ by specifying an inner product
\begin{align}
(c, {c'}) = \delta_{c,c'},
\end{align}
where $c$ and $c'$ are cells of the same dimension.
Between these vector spaces, boundary operators
\begin{align}
\partial_2: C_2 \to C_1,\quad \partial_1:C_1 \to C_0
\end{align}
are defined by taking the oriented boundaries of cells.
These operators satisfy $\partial_1\circ\partial_2 = 0$ since the boundary of a boundary is empty.

For notational consistency with the continuum case, we denote elements of $C_p$ and $C^p$ as $p$-forms.
We rewrite the boundary operators as
\begin{align}
\delta_2 \coloneqq \partial_2,\quad \delta_1 \coloneqq \partial_1,
\end{align}
and call them codifferentials.
We also define the exterior derivatives as $d_i = \delta_{i+1}^\mathrm{T}$ by taking the transpose in the basis $\{c\}$.
These operators satisfy $d_1\circ d_0 = 0$.

In the case of the two-dimensional square lattice, the explicit actions of these operators $d_0$, $d_1$, $\delta_1$, and $\delta_2$ are given by
\begin{align}&
 d_0v = d_0\Img[0.2]{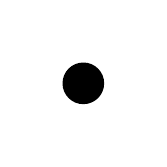} = \Img[0.2]{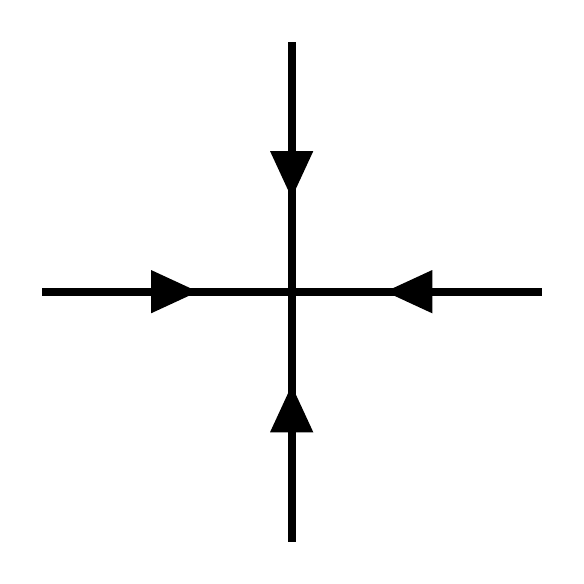}, \\ &
 d_1e = d_1\Img[0.2]{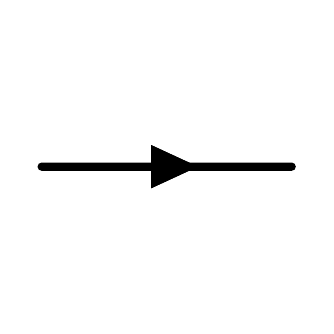} = \Img[0.2]{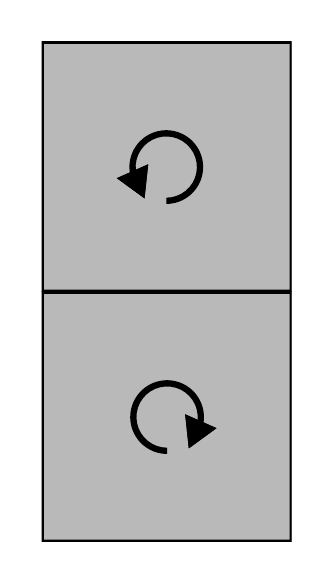}, \\ &
 \delta_1e = \delta_1\Img[0.2]{line.pdf} = \Img[0.2]{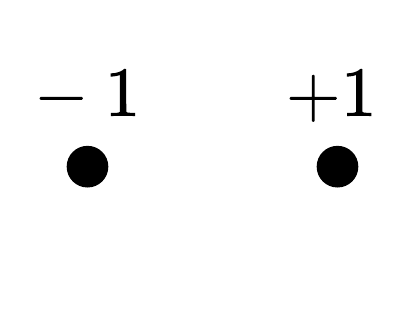},\\ &
 \delta_2f = \delta_2\Img[0.2]{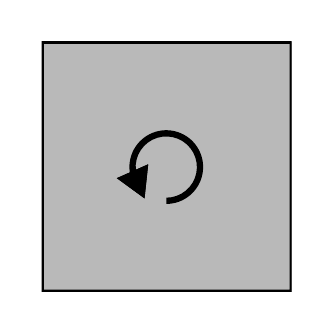} = \Img[0.2]{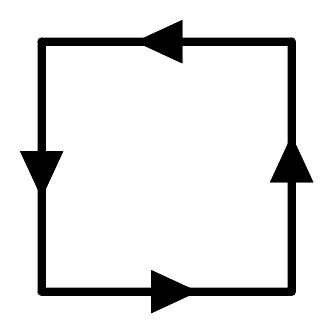}.
\end{align}
In the figures, the arrows and signs indicate the orientations of cells, and opposite orientations give negative signs.
The cells in the same figure are summed with appropriate signs.

The Laplacian operators $\Delta_i: C_i \to C_i$ are defined as
\begin{align}
\Delta_i &= d_{i-1}\delta_i +\delta_{i+1}d_{i}.
\end{align}
These are positive semidefinite operators and correspond to $-\nabla^2$ in the continuum limit.

The Hodge decomposition theorem holds in this discrete setting as well.
The $p$-form space $C_p$ can be decomposed as
\begin{align}
C_p = \Im d_{p-1} \oplus \Im \delta_{p+1} \oplus \ker \Delta_p.
\end{align}
Elements of each subspace are called exact forms, coexact forms, and harmonic forms, respectively.
The number of independent harmonic $p$-forms coincides with the number of $p$-dimensional nontrivial cycles on the lattice.

\subsection{Lattice QBT Model}

Let us consider lattice free-fermion models with QBT.
The construction is parallel to that of the continuum model since we are using consistent notations in both cases.
Since we do not specify the lattice structure, we work directly in real space and do not use momentum-space representations.

I assign fermions to all edges and denote their creation and annihilation operators as $\hat{\psi}^\dagger(e)$ and $\hat{\psi}(e)$, respectively.
These satisfy the canonical anticommutation relations
\begin{align}
\{\hat{\psi}(e), \hat{\psi}^\dagger(e')\} = \lp e,{e'} \rp = \delta_{e,e'}.
\end{align}
I also define the 1-form fermion operators as
\begin{align}
\hat{\psi} \coloneqq \sum_{e \in E} \hat{\psi}(e){e},\quad
\hat{\psi}^\dagger \coloneqq \sum_{e \in E} \hat{\psi}^\dagger(e){e}.
\end{align}

The Hamiltonian is given by
\begin{align}
\hat{H} &
= t_+\lp d\hat{\psi}^\dagger, d\hat{\psi} \rp + t_-\lp \delta\hat{\psi}, \delta\hat{\psi}^\dagger \rp\nonumber\\ &
= t_+\sum_{f \in F} d\hat{\psi}^\dagger(f)d\hat{\psi}(f) + t_-\sum_{v \in V} \delta\hat{\psi}(v)\delta\hat{\psi}^\dagger(v)\\ &
= \lp \hat{\psi}^\dagger,(t_+\delta d-t_-d\delta)\hat{\psi} \rp + \mathrm{const.},
\label{discrete Hamiltonian}
\end{align}
where $t_{\pm}$ are positive parameters and I have omitted the subscripts for $d$ and $\delta$.
This Hamiltonian is constructed from two types of local operators $d\hat{\psi}(f)$ and $\delta\hat{\psi}^\dagger(v)$.
These operators satisfy the anticommutation relations
\begin{align}&
\{d\hat{\psi}(f), \delta\hat{\psi}^\dagger(v)\} = \lp \delta f, dv \rp = \lp f, d^2v \rp = 0.
\end{align}
I note that analogous relations hold in general frustration-free free-fermion systems~\cite{onoFrustrationfreeFreeFermions2025,masaokaFrustrationfreeFreeFermions2025}.

The energy levels are obtained by diagonalizing the matrix $t_+\delta_2d_1-t_-d_0\delta_1$.
Recalling the Hodge decomposition $C_1 = \Im d_0 \oplus \Im \delta_2 \oplus \ker \Delta_1$, the first term $t_+\delta_2d_1$ acts only on the coexact part $\Im \delta_2$ while the second term $-t_-d_0\delta_1$ acts only on the exact part $\Im d_0$.
Therefore, the one-particle Hamiltonian matrix is block-diagonalized as
\begin{align}
H = \begin{pmatrix}
-t_-d_0\delta_1 & 0 & 0 \\
0 & t_+\delta_2d_1 & 0 \\
0 & 0 & 0
\end{pmatrix}.
\end{align}
Since both $d_0\delta_1$ and $\delta_2d_1$ are positive semidefinite, $\Im d_0$ gives negative energy modes and $\Im \delta_2$ gives positive energy modes.
Harmonic 1-forms $h^a_1 \in \ker \Delta_1$ represent zero modes, and their creation and annihilation operators are given by $\lp h^a_1, \hat{\psi}^\dagger \rp$ and $\lp h^a_1,\hat{\psi} \rp$.
The ground-state degeneracy arises from the freedom of choosing whether each zero mode is occupied or not.

The fermion creation operator for the orbital $dv$ corresponding to a vertex $v \in V$ is given by
\begin{align}
\lp dv,\hat{\psi}^\dagger \rp = \lp v,\delta\hat{\psi}^\dagger \rp = \delta\hat{\psi}^\dagger(v).
\end{align}
Therefore, the ground state, as the state with all negative energy modes filled, might be expressed as
\begin{align}
| \mathrm{GS} \rangle \overset{?}{=} \prod_{v \in V} \delta\hat{\psi}^\dagger(v) | 0 \rangle.
\label{lattice wrong GS}
\end{align}
However, since $\delta\hat{\psi}^\dagger(v)$ satisfies the relation $\lp h_0, \delta\hat{\psi}^\dagger \rp = \lp dh_0, \hat{\psi}^\dagger \rp = 0$ for any harmonic $0$-form $h_0 \in \ker \Delta_0$, the operators $\{\delta\hat{\psi}^\dagger(v)\}_{v\in V}$ are linearly dependent, and the harmonic modes $\lp h_0, \delta\hat{\psi}^\dagger \rp $ must be removed.
We assume there is only one harmonic $0$-form since we are interested in connected lattices.

The state in Eq.~\eqref{lattice wrong GS} can be expressed using a fermionic path integral as
\begin{align} &
\prod_{v \in V} \delta\hat{\psi}^\dagger(v) | 0 \rangle
\propto \prod_{v\in V}\left[-g\int\exp\left(\theta(v)\delta\hat{\psi}^\dagger(v)\right)\lvec{\mathrm{d}}\theta(v)\right]| 0 \rangle
\end{align}
where $\theta(v)$ is a Grassmann variable defined on a vertex $v \in V$.
I have also introduced a constant $g = 1/\sqrt{4\pi}$ for later convenience.

To remove harmonic modes, I modify the path integral as
\begin{align}
| \xi \rangle
&
\coloneqq \frac{1}{\sqrt{Z}}\int\lp h_0, \theta \rp\exp\left(-g\sum_{v\in V}\theta(v)\delta\hat{\psi}^\dagger(v)\right)| 0 \rangle \lvec{\mathcal{D}}\theta \nonumber\\ &
= \frac{1}{\sqrt{Z}}\int\lp h_0, \theta \rp\exp(-g\lp \theta, \delta\hat{\psi}^\dagger \rp)| 0 \rangle \lvec{\mathcal{D}}\theta \nonumber\\ &
= \frac{1}{\sqrt{Z}}\int\lp h_0, \theta \rp\exp(-g\lp d\theta, \hat{\psi}^\dagger \rp)| 0 \rangle \lvec{\mathcal{D}}\theta \nonumber\\ &
= \frac{1}{\sqrt{Z}}\int\xi| gd\theta \rangle_\mathrm{fc}\lvec{\mathcal{D}}\theta.
\label{lattice GS}
\end{align}
Here, $\sqrt{Z}$ is a normalization constant, $\xi \coloneqq \lp h_0, \theta \rp$, and $| gd\theta \rangle_\mathrm{fc}$ is a fermionic coherent state.

Thus, we end up with the same expression as that of the continuum theory in Eq.~\eqref{path integral representation of GS}.
The remaining steps are all identical to those in the continuum case.
The ground states satisfy the frustration-free conditions
\begin{align}
d\hat{\psi}(f)| \mathrm{GS} \rangle = 0,\quad
\delta\hat{\psi}^\dagger(v)| \mathrm{GS} \rangle = 0.
\end{align}
The action of the discrete symplectic fermion theory is given by
\begin{align}
S[\theta, \theta^*] = -g^2\lp d\theta^*, d\theta \rp = \frac1{4\pi}\sum_{e \in E}d{\theta}(e) d\theta^*(e).
\end{align}
The correspondence between the fermion operators and the fields of symplectic fermion theory is given by
\begin{align}&
\hat{\psi}(e) \leftrightarrow \frac{d\theta(e)}{\sqrt{4\pi}},\quad
\hat{\psi}^\dagger(e) \leftrightarrow \frac{d\theta^*(e)}{\sqrt{4\pi}}.
\end{align}

\subsection{Tight-binding model on line graphs}

The lattice QBT model introduced above admits a natural reinterpretation as a tight-binding model defined on the line graph of the original lattice.
Line graphs are constructed by replacing each edge of the original lattice with a vertex, and by connecting two such vertices whenever the corresponding edges share a common endpoint.
The line graph of the square lattice is known as the checkerboard lattice and that of the honeycomb lattice is known as the kagome lattice.

In this correspondence between lattices, a fermion originally assigned to an edge is reinterpreted as a fermion living on the associated vertex of the line graph.
To make this identification precise, we must fix an orientation for every edge in the original lattice.
For each edge $e$, we choose one of the two possible orientations $\hat{\psi}(e)$ and $-\hat{\psi}(e)$ as the reference orientation that defines the corresponding vertex operator on the line graph.

\begin{figure}[t]
    \centering
    \includegraphics[width=0.9\hsize]{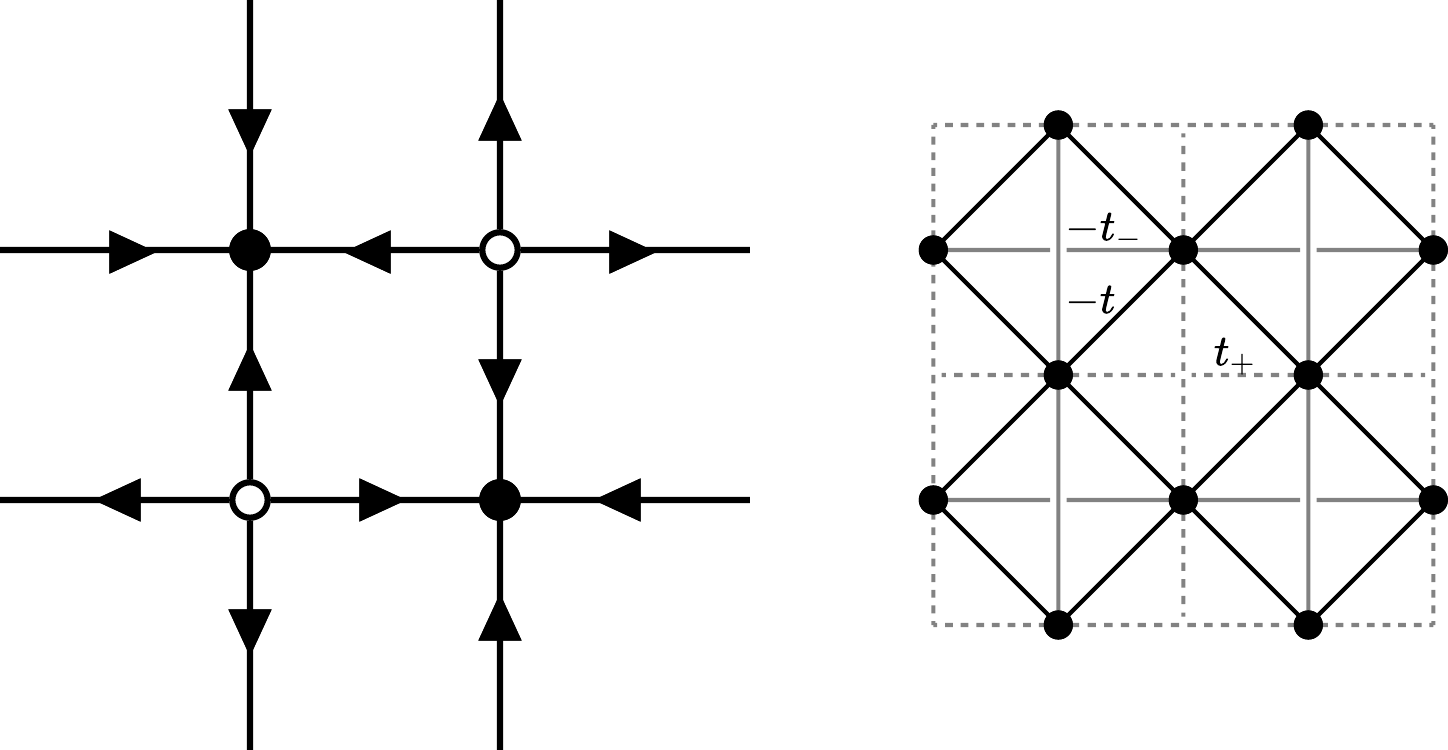}
    \label{square and checkerboard}
    \caption{A square lattice (left) and a checkerboard lattice (right). Edges of the square lattice in the staggered orientation correspond to the vertices of the checkerboard lattice. The hopping coefficients for the Hamiltonian~\eqref{discrete Hamiltonian} are $-t_-$ along the horizontal and vertical full lines, $t_+$ along the horizontal and vertical dashed lines, and $-t = -(t_++t_-)$ along the diagonal lines.}
\end{figure}

Uniform hopping on the line graph requires a coherent choice of edge orientations.
For the square and checkerboard lattices, this can be achieved by adopting a staggered pattern of orientations, illustrated in Fig.~\ref{square and checkerboard}.
With this choice, hopping amplitudes corresponding to Eq.~\eqref{discrete Hamiltonian} for vertical and horizontal edges alternate between $-t_-$ and $t_+$ depending on sublattices, while diagonal hoppings consistently have amplitude $-t \coloneqq -(t_+ + t_-)$.

In this setup, the tight-binding Hamiltonian has a band touching point at $\bm{k} = (\pi, \pi)$ because of the staggered orientations.
If either of the side lengths of the lattice is odd, there is no zero mode since there is no mode with $\bm{k} = (\pi,\pi)$.
In terms of the original square lattice model, this corresponds to imposing antiperiodic boundary conditions along the direction(s) with odd length, thereby removing the harmonic 1-forms.

\section{Short review on symplectic fermion theory}
\label{sec: review of symplectic fermion}

In this section, we briefly review two-dimensional symplectic fermion theory, focusing on its basic aspects as a logarithmic conformal field theory.
We restrict ourselves to those elements that will be needed later.
For more comprehensive discussions of symplectic fermions and logarithmic CFTs, see Refs.~\cite{gurarieLogarithmicOperatorsConformal1993,    kauschCuriositiesC21995,kauschSymplecticFermions2000,flohrBitsPiecesLogarithmic2003,creutzigLogarithmicConformalField2013}.

The two-dimensional symplectic fermion CFT is one of the most well-understood examples of logarithmic CFT.
The universality class of this theory (and its relatives sharing the same central charge) appears in various contexts of statistical mechanics, such as the critical dense polymers~\cite{saleurPolymersPercolationTwo1992,ivashkevichCorrelationFunctionsDense1999}, the abelian sandpile model~\cite{dharSelforganizedCriticalState1990,ruelleLogarithmicConformalInvariance2013} and specific dimer models~\cite{izmailianLogarithmicConformalField2005,adame-carrilloDiscreteSymplecticFermions2025}.
In the quantum realm, it describes the Haldane-Rezayi state in quantum Hall systems~\cite{haldaneSpinsingletWaveFunction1988,gurarieHaldaneRezayiQuantumHall1997}.
It also describes non-Hermitian critical free-fermion systems~\cite{changEntanglementSpectrumEntropy2020}.

The action of the symplectic fermion theory is given by
\begin{align}
S[\theta,\theta^*]
&
= \frac1{4\pi} \int\mathrm{d}^2\bm{x}\,\partial_i\theta(\bm{x})\partial^i\theta^*(\bm{x})
\nonumber\\
&
= \frac1{\pi}\int\mathrm{d}^2z\,\partial\theta(z,\bar{z}) \bar\partial\theta^*(z,\bar{z}) \nonumber\\
&
= \frac1{2\pi}\int\mathrm{d}^2z \,\varepsilon_{\alpha\beta}\partial\theta^{\alpha}(z,\bar{z})\bar{\partial}\theta^\beta(z,\bar{z}),
\label{action of SF}
\end{align}
where $(\theta^1, \theta^2) = (\theta, \theta^*)$ are Grassmann-valued fields and $\varepsilon$ is the symplectic form with $\varepsilon_{12} = -\varepsilon_{21} = 1$.
In the final expression, the symplectic symmetry of the theory becomes explicit.
The action is invariant under the transformation $\theta^\alpha \mapsto {U_\alpha}^\beta\theta^\beta$ with $U \in \mathrm{SL}(2,\mathbb{C}) = \mathrm{Sp}(2,\mathbb{C})$.

\subsection{Correlation functions}

The correlation functions of this theory are defined by the path integral
\begin{align}
 \langle X \rangle = \frac{1}{Z}\int\mathcal{D}\theta\mathcal{D}\theta^*X e^{-S[\theta,\theta^*]},
\end{align}
where $X$ is an arbitrary operator and $Z$ is the partition function defined later.
Since the action is quadratic, correlation functions can be computed exactly using Gaussian integration.

Let us first consider the expectation value of the identity operator.
Because the action does not contain the zero modes explicitly, the Grassmann integral over the zero modes vanishes unless they are saturated by operator insertions.
As a result, the vacuum expectation value of the identity operator vanishes, $\langle \mathbb{I} \rangle = 0$.
Such zeros in correlation functions of primary fields are a general feature of logarithmic conformal field theories.
Since the normalization of correlation functions is not fixed by $\langle \mathbb{I} \rangle = 1$, we instead set $\langle \xi^*\xi \rangle = 1$ as the normalization condition, where $\xi$ and $\xi^*$ are the zero modes defined by
\begin{align}
 \xi \coloneqq \frac1{\mathcal{V}}\int\mathrm{d}^2z\,\theta^1(z,\bar z),\quad
 \xi^* \coloneqq \frac1{\mathcal{V}}\int\mathrm{d}^2z\,\theta^2(z,\bar z),
\end{align}
and $\mathcal{V}$ is the volume of the space.
Equivalently, we define the partition function by
\begin{align}
Z \coloneqq \int\mathcal{D}\theta\mathcal{D}\theta^*\xi^*\xi e^{-S[\theta,\theta^*]}.
\end{align}

With this convention, the simplest non-vanishing correlation function is
\begin{align}
\langle \theta^*(z,\bar{z})\theta(w,\bar{w}) \rangle
&
= \langle \xi^*\xi \rangle = 1,
\end{align}
which simply reflects the chosen normalization of the zero modes.
Including the zero-mode insertions, the two-point function of $\theta^\alpha$ is calculated as
\begin{align}
 \langle \xi^*\xi\theta^\alpha(z,\bar{z})\theta^\beta(w,\bar{w}) \rangle
 = \varepsilon^{\alpha\beta}\ln|z-w|^2,
\end{align}
where $\varepsilon^{\alpha\beta} = -\varepsilon_{\alpha\beta}$.
For higher-point functions, we obtain
\begin{align}
 \langle \xi^*\xi\theta^{\alpha_1}(z_1,\bar{z}_1)\cdots\theta^{\alpha_n}(z_n,\bar{z}_n) \rangle \nonumber\\
 = \operatorname{Pf} \left( \varepsilon^{\alpha_i\alpha_j}\ln|z_i-z_j|^2 \right)_{1\le i,j\le n}.
\end{align}
By differentiating these correlation functions, we obtain those for $\partial\theta^\alpha$ as
\begin{align}
 \langle \xi^*\xi\partial\theta^{\alpha_1}(z_1)\cdots\partial\theta^{\alpha_n}(z_n) \rangle 
 = \operatorname{Pf} \left( \frac{\varepsilon^{\alpha_i\alpha_j}}{(z_i-z_j)^2} \right)_{1\le i,j\le n},
\end{align}
which reproduce the correlation functions in the QBT system.

\subsection{Chiral sector}

The classical equation of motion derived from Eq.~\eqref{action of SF} is given by
\begin{align} 
\frac{\delta S}{\delta\theta^\alpha(z,\bar z)} = -\frac1{\pi}\varepsilon_{\alpha\beta}\partial\bar\partial\theta^\beta(z,\bar z) = 0,
\end{align}
which implies that $\theta^\alpha(z,\bar z)$ decomposes as
\begin{align}
 \theta^\alpha(z,\bar z) = \theta^\alpha(z) + \bar{\theta}^\alpha(\bar z). \label{decomposition of theta}
\end{align}
Here, there remains a freedom in the assignment of the zero modes $\xi^\alpha$, and we assign $\xi^\alpha/2$ to each sector.
Do not confuse the anti-holomorphic fields $\bar{\theta}^\alpha(\bar z)$ with $\theta^*(z,\bar{z})$.

We also have the Schwinger-Dyson equation
\begin{align}
 \theta^\alpha(z,\bar z)\partial\bar\partial{\theta}^{\beta}(w,\bar w) = \pi\varepsilon^{\alpha\beta}\delta^2(z-w,\bar z-\bar w)\mathbb{I},
\end{align}
where $\varepsilon^{\alpha\beta} = -\varepsilon_{\alpha\beta}$ and $\mathbb{I}$ is the identity operator.
Using the identity $\partial\bar\partial\ln|z-w|^2 = \pi\delta^2(z-w,\bar z-\bar w)$, we can rewrite this as
\begin{align}
 \theta^\alpha(z,\bar z)\theta^\beta(w,\bar w) \sim \varepsilon^{\alpha\beta}\ln|z-w|^2 \mathbb{I},
    \label{OPE of theta}
\end{align}
where $\sim$ means that non-singular terms are omitted.
For the holomorphic and anti-holomorphic sectors, we have
\begin{align}&
 \theta^\alpha(z)\theta^\beta(w) \sim \varepsilon^{\alpha\beta}\ln(z-w)\mathbb{I}, \label{OPE of holo theta}\\ 
 &
\bar{\theta}^\alpha(\bar z)\bar{\theta}^\beta(\bar w) \sim \varepsilon^{\alpha\beta}\ln(\bar z-\bar w)\mathbb{I}, \label{OPE of anti-holo theta} \\ 
&
 \theta^\alpha(z)\bar{\theta}^\beta(\bar w) \sim 0. \label{independence of two sectors}
\end{align}

The energy-momentum tensor of this theory is given by
\begin{align}
 T(z) &= -2\pi T_{zz} = {:}\partial\theta^*\partial\theta{:}(z),
\end{align}
and similarly for the anti-holomorphic part.
The operator product expansion of the two energy-momentum tensors is calculated using Eq.~\eqref{OPE of holo theta} as
\begin{align}
 T(z)T(w)
 \sim \frac{-\mathbb{I}}{(z-w)^4} + \frac{2T(w)}{(z-w)^2} + \frac{\partial T(w)}{z-w}.
\end{align}
From the coefficient of the first term, we conclude that the central charge of this theory is $c = -2$.

The fermionic fields $\theta^\alpha(z)$ are primary fields with conformal weight $h = 0$ since their operator product expansions with $T(z)$ are given by
\begin{align}
 T(z)\theta^\alpha(w)
 \sim \frac{\partial\theta^\alpha(w)}{z-w}.
\end{align}
Differentiating this, we find
\begin{align}
 T(z)\partial\theta^\alpha(w)
 \sim \frac{\partial\theta^\alpha(w)}{(z-w)^2} + \frac{\partial^2\theta^\alpha(w)}{z-w}.
\end{align}
Thus, $\partial\theta^\alpha(z) \eqcolon \psi^\alpha(z)$ are primary fields with conformal weight $h = 1$.

Another field with $h = 0$ is defined as
\begin{align}
 \omega(z) &
 \coloneqq {:}\theta\theta^*{:}(z) \nonumber\\
    &
 = \lim_{w\to z}\left( \theta(z)\theta^*(w) + \ln(z-w)\mathbb{I} \right).
\end{align}
The operator product expansion of $T(z)$ and $\omega(w)$ is given by
\begin{align}
 T(z)\omega(w)
 = \frac{\mathbb{I}}{(z-w)^2}
    + \frac{\partial\omega(w)}{z-w},
    \label{log OPE}
\end{align}
which implies that the field $\omega(z)$ is a logarithmic partner of the identity operator $\mathbb{I}$.
The Virasoro mode $L_0$ acts on $| 0 \rangle$ and $| \omega \rangle = \omega(0)| 0 \rangle $ as
\begin{align}
 L_0|\omega\rangle = |0\rangle,\quad L_0|0\rangle = 0,
\end{align}
which shows a Jordan block structure.
Exponentiating $L_0$, we find
\begin{align}
 e^{\alpha L_0}|\omega\rangle = |\omega\rangle + \alpha|0\rangle,\quad e^{\alpha L_0}|0\rangle = |0\rangle.
\end{align}
In particular, under the $2\pi$ rotation $e^{2\pi i L_0}$, the state $| \omega \rangle$ acquires an additional term $2\pi i | 0 \rangle$.

\section{Anyonic excitations}
\label{sec: anyons}

In this section, I discuss excitations with localized energy density in the QBT system.
I observe three kinds of excitations, which I call Dirichlet, Neumann, and composite excitations.
While the QBT system is gapless, these excitations exhibit properties similar to anyons in topological phases.
Moving these excitations along non-contractible loops induces transitions between degenerate ground states.
I also discuss the braiding and spin of these excitations.

\subsection{Construction of anyonic excitations}
\subsubsection{Dirichlet excitations}

Since the $\theta$ and $\theta^*$ fields always appear differentiated in the correspondence of Eq.~\eqref{correspondence for QBT}, their logarithmic correlations seem not to be observed in physical quantities.
Nevertheless, excitations corresponding to $\theta$ field can be observed as anyonic excitations. 

Let us consider local excitations where the state is allowed to have non-zero energy density only at localized points. 
Such local excitations can be obtained as additional ground states when some local terms are removed from the Hamiltonian.
In the continuum theory, this setting corresponds to creating punctures in the spatial manifold.


In this subsection, I assume that the spatial manifold is spherical, eliminating the possibility of topological degeneracy.
Let us first consider the Hamiltonian where the local term $\delta\hat{\psi}(\bm{x})\delta\hat{\psi}^\dagger(\bm{x})$ is removed at point $\bm{x}_1$:
\begin{align}
\hat{H} = t_+ \sum_{\tilde{\bm{x}} \in F} d\hat{\psi}^\dagger(\tilde{\bm{x}})d\hat{\psi}(\tilde{\bm{x}}) + t_-\sum_{\substack{\bm{x}\in V\\\bm{x} \ne \bm{x}_1}}\delta\hat{\psi}(\bm{x})\delta\hat{\psi}^\dagger(\bm{x}).
\end{align}
For this modified Hamiltonian, the frustration-free conditions imposed on ground states are
\begin{align}
d\hat{\psi}(\bm{x})| \mathrm{GS} \rangle = 0,\quad
\delta\hat{\psi}^\dagger(\bm{x})| \mathrm{GS} \rangle = 0~(\bm{x}\ne\bm{x}_1).
\end{align}
However, dropping one condition does not increase the ground state degeneracy since $\{\delta\hat{\psi}^\dagger(\bm{x})\}$ are linearly dependent and satisfy
\begin{align}
 \delta\hat{\psi}^\dagger(\bm{x}_1) = -\sum_{\bm{x}\ne\bm{x}_1}\frac{h_0(\bm{x})}{h_0(\bm{x}_1)}\delta\hat{\psi}^\dagger(\bm{x}),\quad h_0\in\ker\Delta_0.
\end{align}

Next, let us consider the case where the local term $\delta\hat{\psi}(\bm{x})\delta\hat{\psi}^\dagger(\bm{x})$ is removed at two points $\bm{x}_1, \bm{x}_2$. In this case, the additional ground state is given by
\begin{align}
| \theta(\bm{x}_1)\theta(\bm{x}_2) \rangle
&
\coloneqq \frac{1}{\sqrt{Z}}\int\theta(\bm{x}_1)\theta(\bm{x}_2) | gd\theta \rangle_\mathrm{fc}\lvec{\mathcal{D}}\theta.
\end{align}
Since $\theta(\bm{x})$ is equivalent to the delta function $\delta(\theta(\bm{x}))$, this state represents two punctures with the Dirichlet boundary condition at $\bm{x}_1$ and $\bm{x}_2$.
For the original Hamiltonian, this state becomes an excited state with localized energy density at $\bm{x}_1$ and $\bm{x}_2$. Note that we are considering a gapless system, and there exist arbitrarily many energy levels with lower energy than this state. We call these local excitations Dirichlet excitations or $\theta$ excitations.

The Dirichlet excitations, despite having localized energy density, are not created by local operators like $\hat{O}(\bm{x}_1)\hat{O}'(\bm{x}_2)$ from the ground state $| \xi \rangle$. Instead, they are created by non-local string operators.
This can be seen as
\begin{align}
 | \theta(\bm{x}_1)\theta(\bm{x}_2) \rangle
    &
 = \frac1{\sqrt{Z}}\int(\theta(\bm{x}_1) - \theta(\bm{x}_2))\xi| gd\theta \rangle_\mathrm{fc}\lvec{\mathcal{D}}\theta \nonumber\\ &
 = \frac1{\sqrt{Z}}\int\int_{\bm{x}_2}^{\bm{x}_1}\mathrm{d}\theta\xi| gd\theta \rangle_\mathrm{fc}\lvec{\mathcal{D}}\theta \nonumber\\ &
 = \sqrt{4\pi} \int_{\bm{x}_2}^{\bm{x}_1}\hat{\psi}| \xi \rangle.
\end{align}
The curve connecting $\bm{x}_1$ and $\bm{x}_2$ can be continuously deformed since $d\hat{\psi}(\bm{x}) | \xi \rangle = 0$.

\begin{figure}[t]
    \centering
    \includegraphics[width=0.9\hsize]{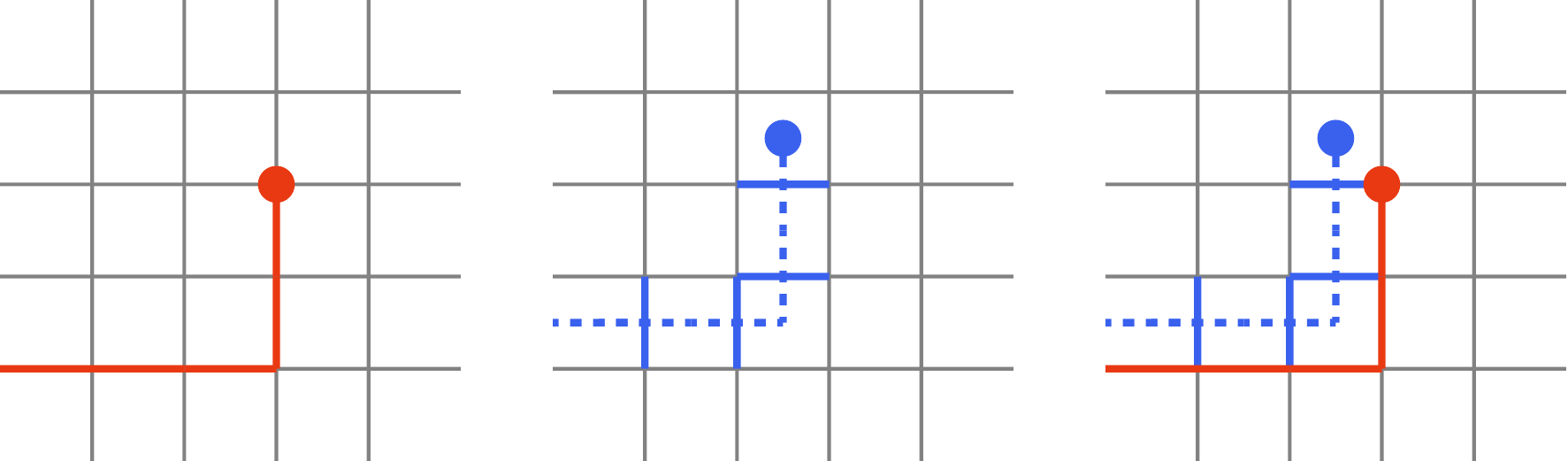}
    \caption{Realizations of the Dirichlet ($\theta$) excitation (left), Neumann ($\phi^*$) excitation (center), and the composite ($\phi^*\theta$) excitations (right) in a square lattice.
 The Dirichlet excitations are placed on vertices, and the Neumann excitations are placed on faces.}
    \label{fig: anyons}
\end{figure}

In a similar manner, we can create multiple Dirichlet excitations by inserting $\theta$ fields in the path integral.
For $n$ points $\bm{x}_1, \ldots, \bm{x}_n$, the excited state is given by
\begin{align} &
| \theta(\bm{x}_1)\cdots\theta(\bm{x}_n) \rangle \nonumber\\ &
\coloneqq \frac1{\sqrt{Z}}\int\mathcal{D}\theta\,\theta_1\cdots\theta_n | gd\theta \rangle_\mathrm{fc} \nonumber\\ &
= \frac1{\sqrt{Z}}\int (\theta_1-\theta_2)(\theta_1-\theta_3)\cdots(\theta_1-\theta_{n})\xi| gd\theta \rangle_\mathrm{fc}\lvec{\mathcal{D}}\theta \nonumber\\ &
= (\sqrt{4\pi})^n \int_{\bm{x}_2}^{\bm{x}_1}\hat{\psi}\int_{\bm{x}_3}^{\bm{x}_1}\hat{\psi}\cdots\int_{\bm{x}_n}^{\bm{x}_1}\hat{\psi} | \xi \rangle,
\end{align}
where $\theta_i \coloneqq \theta(\bm{x}_i)$. 
On the other hand, excitations of the form
\begin{align} &
 | \xi\theta(\bm{x}_1)\cdots\theta(\bm{x}_n) \rangle \nonumber\\ &
 \coloneqq \frac1{\sqrt{Z}}\int\xi \theta(\bm{x}_1)\cdots\theta(\bm{x}_n) | gd\theta \rangle_\mathrm{fc}\lvec{\mathcal{D}}\theta \nonumber\\ &
 = \frac{1}{\sqrt{Z}\mathcal{V}}\int\mathrm{d}^2\bm{x}\int\xi \theta(\bm{x}) \theta(\bm{x}_1)\cdots\theta(\bm{x}_n) | gd\theta \rangle_\mathrm{fc}\lvec{\mathcal{D}}\theta
\end{align}
do not have localized energy density.
In the lattice model, the Dirichlet excitations are placed on vertices as depicted in Fig.~\ref{fig: anyons}.

Let us denote $| \theta(\bm{x})\theta(\bm{y}) \rangle^\dagger = \langle \theta^*(\bm{x})\theta^*(\bm{y}) |$.
The overlap of Dirichlet excitations at different positions is calculated as
\begin{align}&
\langle \theta^*(\bm{x}_1)\theta^*(\bm{x}_2) | \theta(\bm{y}_1)\theta(\bm{y}_2) \rangle \nonumber\\ &
= \langle \bar\xi\xi(\theta^*(\bm{x}_2)-\theta^*(\bm{x}_1))(\theta(\bm{y}_2)-\theta(\bm{y}_1)) \rangle \nonumber\\
&
= \ln \frac{|\bm{x}_1-\bm{y}_2|^2|\bm{x}_2-\bm{y}_1|^2}{|\bm{x}_1-\bm{y}_1|^2|\bm{x}_2-\bm{y}_2|^2}
\label{four point correlation function}
\end{align}
This looks weird since it does not decay with distance.
This is because the state $| \theta(\bm{x}_1)\theta(\bm{x}_2) \rangle$ is not normalized.
The norm of $| \theta(\bm{x}_1)\theta(\bm{x}_2) \rangle$ is given using Eq.~\eqref{four point correlation function} as
\begin{align}
\|| \theta(\bm{x}_1)\theta(\bm{x}_2) \rangle\| = \ln\frac{|\bm{x}_1-\bm{x}_2|^2}{a^2},
\end{align}
where $a$ is the short-distance cutoff.
The normalized correlation function is given by
\begin{align}&
\frac{\langle \theta^*(\bm{x}_1)\theta^*(\bm{x}_2) | \theta(\bm{y}_1)\theta(\bm{y}_2) \rangle}{\|\langle \theta^*(\bm{x}_1)\theta^*(\bm{x}_2)|\|\cdot\|| \theta(\bm{y}_1)\theta(\bm{y}_2) \rangle\|} \nonumber\\ &
= \frac1{2}\ln \frac{|\bm{x}_1-\bm{y}_2||\bm{x}_2-\bm{y}_1|}{|\bm{x}_1-\bm{y}_1||\bm{x}_2-\bm{y}_2|}
\left(\ln\frac{|\bm{x}_1-\bm{x}_2|}{a}\ln\frac{|\bm{y}_1-\bm{y}_2|}{a}\right)^{-\frac1{2}}.
\end{align}
This scales as $O(1/\ln(L/a))$ when $\bm{x}_i-\bm{x}_j = O(L)$.
Thus, the inner product of Dirichlet excitations exhibits logarithmic spatial dependence originating from the correlation function of $\theta(\bm{x})$. 

\subsubsection{Neumann excitations}
Let us next consider a modified Hamiltonian in which the local term $d\hat{\psi}^\dagger(\tilde{\bm{x}})d\hat{\psi}(\tilde{\bm{x}})$ is removed at two dual sites $\tilde{\bm{x}}_1$ and $\tilde{\bm{x}}_2$.
The presence of local excitations corresponding to the $\theta$ field suggests that there should also be local excitations corresponding to the $\theta^*$ field.
However, since the ket of the ground state does not depend on $\theta^*$, there are no local excitations directly associated with $\theta^*$.
Instead, we can construct Neumann excitations by
\begin{align}
| \phi^*(\tilde{\bm{x}}_1)\phi^*(\tilde{\bm{x}}_2) \rangle
= \sqrt{4\pi}\int_{\tilde{\bm{x}}_2}^{\tilde{\bm{x}}_1} \star\hat{\psi}^\dagger | \xi \rangle.
\label{def of Neumann excitation}
\end{align}
A lattice realization of this excitation is depicted in Fig.~\ref{fig: anyons}.
I note that if the spatial manifold has non-trivial topology, this definition depends on the choice of curve connecting the dual sites $\tilde{\bm{x}}_1$ and $\tilde{\bm{x}}_2$.

In Eq.~\eqref{def of Neumann excitation}, $\phi^*$ is the dual field of $\theta^*$.
This duality can be understood by exchanging particles and holes and taking Hodge duals in the definition of the ground state.
Previously, we constructed the ground state by starting from the empty vacuum $| 0 \rangle$ and filling modes of $\delta\hat{\psi}^\dagger$, but we can equally well start from $| \tilde{0} \rangle \coloneqq \prod_{e\in E}\hat\psi^\dagger(e)| 0 \rangle$ with all fermions filled and remove electrons by $d\hat{\psi}$.
In this way, by introducing Grassmann variables $\phi^*(\tilde{\bm{x}})$ on the faces of the lattice, we have
\begin{align} &
 |\tilde{\xi}\rangle
 = \frac1{\sqrt{Z}}\int\lp h_2, \phi^* \rp\exp(-g\lp \delta\phi^*, \hat{\psi} \rp)| \tilde{0} \rangle\lvec{\mathcal{D}}\phi^*,\\
    & h_2 \in \ker \Delta_2.
\label{dual GS}
\end{align}
This ground state $| \tilde{\xi} \rangle$ is equivalent to the previous one $| \xi \rangle$ up to occupation of zero modes.
The operator correspondence for the dual field is given by
\begin{align}
    \hat{\psi}^\dagger \leftrightarrow \frac{\delta\phi^*}{\sqrt{4\pi}},\quad
    \hat{\psi} \leftrightarrow \frac{\delta\phi}{\sqrt{4\pi}}.
    \label{dual correspondence}
\end{align}
Comparing this with Eq.~\eqref{correspondence for QBT}, the relation between $\theta,\theta^*$ and $\phi,\phi^*$ is expressed as
\begin{align}
 d\theta^\alpha = \delta\phi^\alpha, \label{duality relation}
\end{align}
where $(\theta^1, \theta^2) = (\theta, \theta^*)$ and $(\phi^1, \phi^2) = (\phi, \phi^*)$.
Using complex coordinates, Eq.~\eqref{duality relation} is written as
\begin{align}
 \partial\theta^\alpha dz + \bar\partial\bar{\theta}^\alpha d\bar z = i\partial\phi^\alpha(z)dz - i \bar\partial\bar{\phi}^\alpha(\bar z)d\bar z,
\end{align}
which implies
\begin{align}
 \phi^\alpha(z, \bar{z}) =  -i\theta^\alpha(z) + i\bar{\theta}^\alpha(\bar{z}),
\end{align}
up to the freedom of additive zero modes.
I fix this freedom by requiring that the zero modes of $\phi^\alpha$ are $\xi^\alpha$.
The duality between $\theta^\alpha$ and $\phi^\alpha$ is analogous to the electric-magnetic duality or T-duality of the free boson CFTs.
Same as the free boson CFTs, exchanging $\theta^\alpha$ and $\phi^\alpha$ fields flips the Dirichlet boundary condition and the Neumann boundary condition.
Therefore, inserting $\phi^*$ fields in the path integral represents punctures with Neumann boundary conditions on $\theta^*$.

\subsubsection{Composite excitations}
Combining Dirichlet and Neumann excitations, we can also construct neutral composite excitations.
Such an excitation is expressed as
\begin{align}
 \phi^*\theta(z,\bar{z}) &
 = (-i\theta^*(z)+i\bar{\theta}^*(\bar{z}))(\theta(z) + \bar{\theta}(\bar{z})) \nonumber\\ &
 = i\theta(z)\theta^*(z) - i\bar{\theta}(\bar{z})\bar{\theta}^*(\bar{z}) \nonumber\\ &
 = i\omega(z) - i\bar{\omega}(\bar{z}),
\end{align}
where $\omega$ and $\bar{\omega}$ are the logarithmic partners of the identity in the holomorphic and anti-holomorphic sectors, respectively.
Same as other excitations, these composite excitations correspond to string operators as
\begin{align}
 \phi^*\theta(\bm{x}_1)\phi^*\theta(\bm{x}_2)
    &
 \leftrightarrow 4\pi\int_{\tilde{\bm{x}}_2}^{\tilde{\bm{x}}_1} \star\hat{\psi}^\dagger\int_{\bm{x}_2}^{\bm{x}_1}\hat{\psi},
\end{align}
where $\tilde{\bm{x}}_i \in F$ denotes the dual site neighboring the site $\bm{x}_i \in V$.
A lattice realization of the composite excitation is depicted in Fig.~\ref{fig: anyons}.

\subsection{Wilson and 't Hooft line operators} 

While the string operators introduced above resemble Wilson lines and 't Hooft lines in gauge theories, there are two clear differences.  
First, the 1-form fields are Grassmann odd in the present setting.  
Second, the string operators are defined as contour integrals rather than as exponentials of such integrals.  

These differences might be accommodated by introducing formal Grassmann parameters associated with the string operators and defining the Wilson and 't Hooft line operators as
\begin{align}
& \hat{W}_{\Gamma}(\eta) \coloneqq e^{\sqrt{4\pi}\eta\int_\Gamma \hat{\psi}} = 1 + \sqrt{4\pi}\eta\int_\Gamma \hat{\psi}. 
\label{Wilson line} \\
& \hat{V}_{\tilde{\Gamma}}(\tilde{\eta}) \coloneqq e^{\sqrt{4\pi}\tilde{\eta}\int_{\tilde{\Gamma}} \star\hat{\psi}^\dagger} = 1 + \sqrt{4\pi}\tilde{\eta}\int_{\tilde{\Gamma}} \star\hat{\psi}^\dagger. 
\label{tHooft line}
\end{align}
Upon integrating out the Grassmann variables $\eta$ and $\tilde{\eta}$, one recovers the original string operators.  

The string operators creating Dirichlet and Neumann excitations satisfy the anticommutation relation
\begin{align}
    \left\{\int_{\Gamma}\hat{\psi}, \int_{\tilde{\Gamma}}\star\hat{\psi}^\dagger \right\}
    = I(\Gamma, \tilde{\Gamma}),
    \label{anticommutation for string operators}
\end{align}
where $I(\Gamma, \tilde{\Gamma})$ is the intersection number.  

These relations can be rewritten in terms of the Wilson and 't Hooft line operators as
\begin{align}
& \hat{W}_{\Gamma}(\eta)\hat{V}_{\tilde{\Gamma}}(\tilde{\eta})
= e^{4\pi I(\Gamma,\tilde{\Gamma})\tilde{\eta}\eta}\hat{V}_{\tilde{\Gamma}}(\tilde{\eta})\hat{W}_{\Gamma}(\eta).
\end{align}
Thus, braiding a Dirichlet excitation with a Neumann excitation yields a factor $e^{4\pi\tilde{\eta}\eta}$ rather than a pure phase factor.  
We also have
\begin{align}
   &
   \hat{W}_{\Gamma_1}(\eta_1)\hat{W}_{\Gamma_2}(\eta_2)
= \hat{W}_{\Gamma_2}(\eta_2)\hat{W}_{\Gamma_1}(\eta_1), \\
& \hat{V}_{\tilde{\Gamma}_1}(\tilde{\eta}_1)\hat{V}_{\tilde{\Gamma}_2}(\tilde{\eta}_2)
= \hat{V}_{\tilde{\Gamma}_2}(\tilde{\eta}_2)\hat{V}_{\tilde{\Gamma}_1}(\tilde{\eta}_1).
\end{align}

\subsection{Topological degeneracy}

\subsubsection{Closed surface cases}
Let us consider the ground state degeneracy on a spatial manifold without boundaries.
Since the system is gapless, we need to be careful in defining the ground state degeneracy.
In the following, I focus on the degeneracy of the exact ground states and ignore excited states with polynomially small energy.

A standard way to determine the ground-state degeneracy is to count the number of zero modes, which correspond to harmonic 1-forms.
Since the number of independent harmonic 1-forms is given by $2g$ for a genus $g$ surface, the ground state degeneracy is $4^{g}$.

However, I present an alternative argument based on anyonic excitations and topological loop operators.
The following argument is analogous to that for the $\mathbb{Z}_2$ toric code model~\cite{kitaevFaulttolerantQuantumComputation2003}, although there are several differences: (i) the system is gapless, (ii) the Hamiltonian is not a sum of commuting projectors, and (iii) the gauge fields are fermionic.

If the spatial manifold has genus $g > 0$, there exist $2g$ independent non-contractible loops.
String operators along such non-contractible loops are defined as
\begin{align}
    \hat{\Psi}_a \coloneqq \oint_{\Gamma_a} \hat{\psi},\quad
    \hat{\Pi}_a \coloneqq \oint_{\tilde{\Gamma}_a} \star\hat{\psi}^\dagger,
\end{align}
where $a = 1,\ldots,2g$ labels these non-contractible loops.
Here, loops $\Gamma_a$ and $\tilde{\Gamma}_b$ are chosen so that they intersect an odd number of times if $a = b$ and an even number of times if $a \ne b$.
Representative curves can be continuously deformed without changing the action of the loop operators on ground states.
Acting with these loop operators corresponds to a process of moving anyons along the loop.
Note that such moving processes are not realized by unitary transformations.
I also note that these non-contractible loop operators were originally introduced in the context of flat bands~\cite{bergmanBandTouchingRealspace2008,rhimClassificationFlatBands2019,rhimSingularFlatBands2021}.

The loop operators satisfy the relations
\begin{align} &
 \{\hat{\Psi}_a, d\hat{\psi}(\tilde{\bm{x}})\} = \{\hat{\Psi}_a, \delta\hat{\psi}^\dagger(\bm{x})\} = 0, \\ &
 \{ \hat{\Pi}_a, d\hat{\psi}(\tilde{\bm{x}})\} = \{\hat{\Pi}_a, \delta\hat{\psi}^\dagger(\bm{x})\} = 0.
\end{align}
Therefore, these operators preserve the frustration-free conditions and map ground states to ground states.
Note that the loop operators do not commute with the Hamiltonian and thus are not symmetries of the system.

The anticommutation relations among the loop operators are calculated as
\begin{align}&
\{ \hat{\Psi}_a, \hat{\Pi}_b \} = \delta_{ab},\\ &
\{ \hat{\Psi}_a, \hat{\Psi}_b \} = \{ \hat{\Pi}_a, \hat{\Pi}_b \} = 0.
\label{anticommutation of loop operators}
\end{align}
Since these operators form a closed algebra acting on the ground state space, the ground state degeneracy is at least $4^{g}$ on a genus $g$ surface.
Furthermore, since $d\hat{\psi}^\dagger(\tilde{\bm{x}})$ and $\delta\hat{\psi}(\bm{x})$ can generate $2^{|V|+|F|-2} = 2^{|E|-2g}$ linearly independent states from each ground state, the total ground state degeneracy is given by $4^{g}$.

Degenerate ground states satisfy
\begin{align}
   \langle \mathrm{GS}_i | \hat{A} | \mathrm{GS}_j \rangle
   = C\delta_{ij} + O(1/L),
\end{align}
for any local operator $\hat{A}$, where $C$ is a constant and $L$ is the linear system size.
Such algebraic distinctions of ground states deviate from the framework of quasi-topological phases~\cite{bondersonQuasiTopologicalPhasesMatter2013}.

\subsubsection{Dirichlet and Neumann boundary conditions}

In the presence of boundaries on the spatial manifold, the ground-state degeneracy can be altered depending on the boundary conditions.

Let us first consider a connected Dirichlet boundary.
On closed manifolds, the frustration-free conditions are generally written as
\begin{align}
 \oint_{\Gamma} \hat{\psi}| \mathrm{GS} \rangle = 0,\quad
 \oint_{\tilde{\Gamma}} \star\hat{\psi}^\dagger| \mathrm{GS} \rangle = 0,
\end{align}
for any contractible loops $\Gamma$ and $\tilde{\Gamma}$.
In the presence of a Dirichlet boundary $B_\mathrm{D}$ with conditions
\begin{align}
 \theta(\bm{y}) - \theta(\bm{x}) = 0,\quad(\bm{x},\bm{y} \in B_\mathrm{D}),
\end{align}
there are additional frustration-free conditions
\begin{align}
\int_{\bm{x}}^{\bm{y}}\hat{\psi}| \mathrm{GS} \rangle = | (\theta(\bm{y})-\theta(\bm{x}))\xi \rangle = 0,
\end{align}
where the integral is taken along the boundary $B_\mathrm{D}$.

If there are $b_\mathrm{D}$ Dirichlet boundaries $B_{\mathrm{D},j}$ with $j=1,\ldots,b_\mathrm{D}$, the string operator
\begin{align}
    \hat{\Psi}_{\mathrm{D},j} \coloneqq \int_{\bm{x}_1}^{\bm{x}_j}\hat{\psi},~
 (j>1, \bm{x}_{i} \in B_{\mathrm{D},i})
\end{align}
anticommutes with all frustration-free conditions and thus creates an additional ground state.
The conjugate operator is defined as
\begin{align}
    \hat{\Pi}_{\mathrm{D},j} \coloneqq \oint_{\tilde{\Gamma}(B_{\mathrm{D},j})} \star\hat{\psi}^\dagger,
\end{align}
where $\tilde{\Gamma}(B_{\mathrm{D},j})$ is a loop surrounding the boundary $B_{\mathrm{D},j}$.
We do not consider a loop operator $\hat{\Pi}_{\mathrm{D},1}$ since $\sum_{j=1}^{b_\mathrm{D}}\tilde{\Gamma}(B_{\mathrm{D},j})$ is contractible.
The operators $\hat{\Psi}_{\mathrm{D},j}$ and $\hat{\Pi}_{\mathrm{D},j}$ ($j=2,\ldots,b_\mathrm{D}$) satisfy the anticommutation relation
\begin{align}
 \{ \hat{\Psi}_{\mathrm{D},i}, \hat{\Pi}_{\mathrm{D},j} \} = \delta_{i,j},\quad
 (i,j = 2,\ldots,b_\mathrm{D}).
\end{align}
Thus, we have $b_\mathrm{D}-1$ pairs of additional creation and annihilation operators acting on the ground-state space, and the ground-state degeneracy is given by $2^{2g+b_\mathrm{D}-1}$.

\begin{figure}[t]
    \centering
    \includegraphics[width=0.6\hsize]{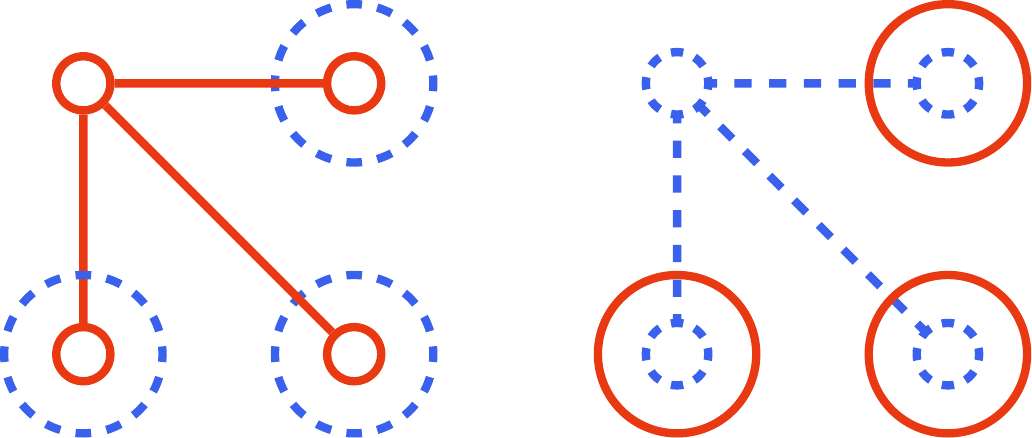}
    \caption{Non-contractible string operators in the presence of Dirichlet and Neumann boundaries.}
    \label{fig: strings with DN}
\end{figure}

On the other hand, when there is a connected boundary with Neumann conditions $B_{\mathrm{N},j}$, we have additional conditions
\begin{align}
 \int_{\tilde{\bm{x}}}^{\tilde{\bm{y}}}\star\hat{\psi}^\dagger| \mathrm{GS} \rangle = 0,\quad
 (\tilde{\bm{x}}, \tilde{\bm{y}} \in B_{\mathrm{N},j}).
\end{align}
Then, we can define additional operators that preserve the frustration-free conditions as
\begin{align}&
    \hat{\Pi}_{\mathrm{N},j} \coloneqq \int_{\tilde{\bm{x}}_1}^{\tilde{\bm{x}}_j}\star\hat{\psi}^\dagger\quad (\tilde{\bm{x}}_i\in B_{\mathrm{N},i}), \\ &
    \hat{\Psi}_{\mathrm{N},j} \coloneqq \oint_{\Gamma(B_{\mathrm{N},j})}\hat{\psi},
\end{align}
where $j = 2,\ldots,b_\mathrm{N}$ and $\Gamma(B_{\mathrm{N},j})$ is a loop surrounding the Neumann boundary $B_{\mathrm{N},j}$.
These operators form additional $b_\mathrm{N}-1$ pairs of creation and annihilation operators.
Thus, if there are $b_\mathrm{D}$ Dirichlet boundaries and $b_\mathrm{N}$ Neumann boundaries, there are $2g + b_\mathrm{D} + b_\mathrm{N} - 2$ non-contractible string operators as illustrated in Fig.~\ref{fig: strings with DN}, and the ground-state degeneracy is given by $2^{2g+b_\mathrm{D}+b_\mathrm{N}-2}$.

\begin{figure}[t]
    \centering
    \includegraphics[width=0.7\hsize]{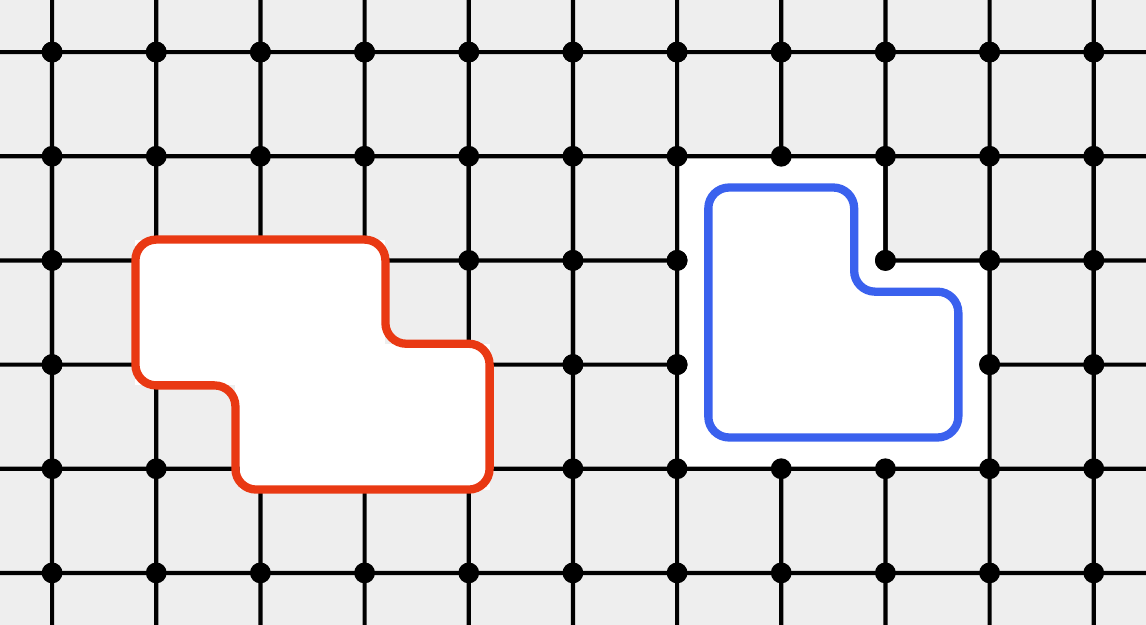}
    \caption{Illustration of a Dirichlet boundary (left) and a Neumann boundary (right) on a square lattice.} 
    \label{fig: boundaries}   
\end{figure}

The lattice realization of Dirichlet and Neumann boundary conditions is depicted in Fig.~\ref{fig: boundaries}.
For the Dirichlet boundary, the fermions on the edges along the boundary are removed so that $d\theta$ vanishes along the boundary.
For the Neumann boundary, the fermions on the edges perpendicular to the boundary are removed so that the normal component of $d\theta^*$ to the boundary vanishes.

\subsection{Non-diagonalizable spin}

Let us see how the anyonic excitations exhibit non-diagonalizable spin.
Similar logarithmic behavior has also been observed in the monodromies of wavefunctions in the Haldane-Rezayi state~\cite{gurarieLogarithmicOperatorsConformal1993}.
However, the construction of the ground states and anyonic excitations considered here is different.

A $2\pi$ rotation is represented by the operator $e^{2\pi i(L_0 - \bar{L}_0)}$.
For the field $\theta(z,\bar{z}) = \theta(z) + \bar{\theta}(\bar{z})$, we have
\begin{align}
   e^{2\pi i(L_0 - \bar{L}_0)}(| \theta \rangle + | \bar{\theta} \rangle)
   &
   = e^{2\pi iL_0}| \theta \rangle + e^{-2\pi i\bar{L}_0}| \bar{\theta} \rangle \nonumber\\ &
   = | \theta \rangle + | \bar{\theta} \rangle,
\end{align}
where we used the fact that $\theta(z)$ and $\bar{\theta}(\bar{z})$ are chiral primary fields with conformal weights $h = 0$ and $\bar{h} = 0$, respectively.
Thus, a single $\theta$ excitation has a trivial spin.
Similarly, a single $\phi^*$ excitation also has a trivial spin.

By contrast, the composite excitation $\phi^*\theta(z,\bar{z})$ exhibits a non-diagonalizable action of the spin operator.
The action of a $2\pi$ rotation on the composite excitation $\phi^*\theta(z,\bar{z}) = i\omega(z) - i\bar{\omega}(\bar{z})$ is given by
\begin{align}&
 e^{2\pi i(L_0 - \bar{L}_0)}(i|\omega\rangle - i|\bar{\omega}\rangle) \nonumber\\ &
 = i(1+2\pi iL_0+\cdots)|\omega\rangle - i(1-2\pi i\bar{L}_0+\cdots)|\bar{\omega}\rangle \nonumber\\ &
 = i(|\omega\rangle + 2\pi i|0\rangle) - i(|\bar{\omega}\rangle - 2\pi i|0\rangle) \nonumber\\ &
 = (i|\omega\rangle - i|\bar{\omega}\rangle) - 4\pi|0\rangle.
\end{align}
Therefore, a $2\pi$ rotation of this anyon produces an additional term proportional to the identity operator, indicating a non-diagonalizable action of the rotation.

\begin{figure}[t]
   \centering
   \includegraphics[width=0.35\hsize]{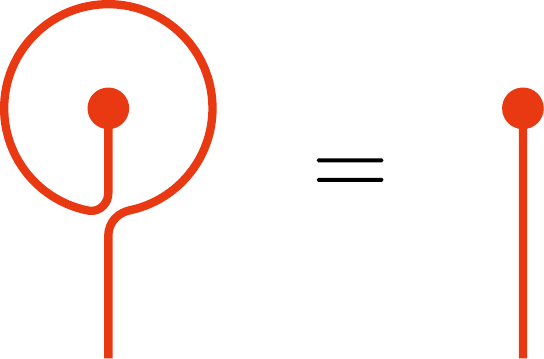}
   \caption{The $2\pi$ rotation of the Dirichlet excitation $\theta(\bm{x})$ is obtained by turning the integration contour around the point $\bm{x}$.}
   \label{fig: trivial spin}
\end{figure}

The same spins can be seen explicitly at the level of ground states and string operators.
First, consider a Dirichlet excitation localized at a point $\bm{x}$, represented by
\begin{align}
   | \theta(\bm{x})\theta(\infty) \rangle = \int_{\infty}^{\bm{x}}\hat{\psi}| \xi \rangle,
\end{align}
where the other endpoint of the string operator is sent to infinity.
For this state, the action of a $2\pi$ rotation is implemented by turning the integration contour around the point $\bm{x}$, as depicted in Fig.~\ref{fig: trivial spin}.
This process produces an additional contour integral, but this contribution vanishes.

\begin{figure}[t]
    \centering
    \includegraphics[width=0.7\hsize]{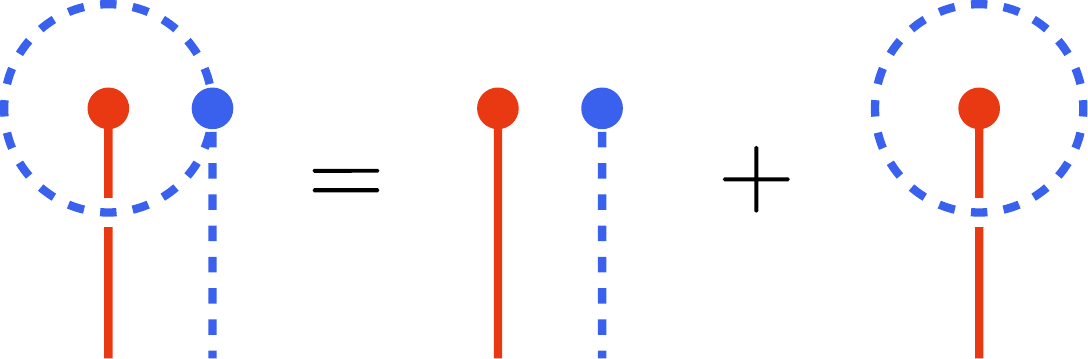}
    \caption{The $2\pi$ rotation of the composite excitation $\phi^*\theta(\bm{x})$ is obtained by winding $\phi^*(\tilde{\bm{x}})$ around $\theta(\bm{x})$.
    This yields an additional contour integral.}
    \label{fig: braiding}
\end{figure}

Next, let us consider a composite excitation localized at a point $\bm{x}$, represented by
\begin{align}
   4\pi \int_{\infty}^{\tilde{\bm{x}}}\star\hat{\psi}^\dagger\int_{\infty}^{\bm{x}}\hat{\psi}| \xi \rangle.
\end{align}
For this state, the action of a $2\pi$ rotation is realized as the braiding between the Dirichlet and Neumann excitations, since both of them individually have trivial spins.
Thus, the $2\pi$ rotation is implemented by winding $\phi^*(\tilde{\bm{x}})$ anticlockwise around $\theta(\bm{x})$, as illustrated in Fig.~\ref{fig: braiding}.

This process yields an additional contour integral,
\begin{align}
    4\pi \oint_{\bm{x}}\star\hat{\psi}^\dagger\int_{\infty}^{\bm{x}}\hat{\psi}| \xi \rangle 
    &= 
    4\pi \left\{\int_{\infty}^{\bm{x}}\hat{\psi}, \oint_{\bm{x}}\star\hat{\psi}^\dagger\right\}| \xi \rangle \nonumber\\ &
 = -4\pi| \xi \rangle,
\end{align}
which is consistent with the previous CFT calculation.

\section{Conclusion}
\label{sec: discussion and outlook}
In this work, I have established that $(d+1)$-dimensional free-fermion models with quadratic band touching realize spatial conformal invariance governed by the $d$-dimensional symplectic fermion theory.
I provided explicit correspondences between operators in the $(d+1)$-dimensional fermionic model and fields in the $d$-dimensional symplectic fermion CFT for both continuum and lattice models.

In particular, in two spatial dimensions I have constructed anyonic excitations that can be understood from the underlying CFT.
These excitations account for the topological ground-state degeneracy and exhibit non-diagonalizable braiding and spin properties.
While these results suggest the existence of more sophisticated frameworks, reminiscent of categorical formulations of topological phases, a more thorough understanding is left for future study.

Finally, I discuss several other future directions.
A straightforward extension of the present work is to incorporate twist fields by introducing $\mathbb{Z}_2$ or $\mathrm{U}(1)$ gauge variables.
Another important direction is to investigate interaction effects at the QBT critical point within the framework of the underlying symplectic fermion CFT and its anyonic excitations.
The correspondence established in this work might be useful for analyzing entanglement properties of QBT systems, which have been investigated in Ref.~\cite{chenScalingEntanglement2+1dimensional2015}.

\section*{Acknowledgements}
I thank Yohei Fuji, Po-Yao Chang, Masahiro Hoshino, Kohei Kawabata, Haruki Watanabe, and Haruki Yagi for fruitful discussions.
I acknowledge support from FoPM, WINGS Program, the
University of Tokyo.

\bibliography{fCQCP}

@article{adame-carrilloDiscreteSymplecticFermions2025,
  title = {Discrete {{Symplectic Fermions}} on {{Double Dimers}} and {{Their Virasoro Representation}}},
  author = {{Adame-Carrillo}, David},
  year = 2025,
  month = mar,
  journal = {Annales Henri Poincar\'e},
  volume = {26},
  number = {3},
  pages = {845--876},
  doi = {10.1007/s00023-024-01455-w},
  langid = {english}
}

@article{anninosHigherSpinRealization2016,
  title = {Higher Spin Realization of the {{DS}}/{{CFT}} Correspondence},
  author = {Anninos, Dionysios and Hartman, Thomas and Strominger, Andrew},
  year = 2016,
  month = feb,
  journal = {Classical and Quantum Gravity},
  volume = {34},
  number = {1},
  pages = {015009},
  publisher = {IOP Publishing},
  doi = {10.1088/1361-6382/34/1/015009},
  langid = {english}
}

@article{ardonneTopologicalOrderConformal2004,
  title = {Topological Order and Conformal Quantum Critical Points},
  author = {Ardonne, Eddy and Fendley, Paul and Fradkin, Eduardo},
  year = 2004,
  month = apr,
  journal = {Annals of Physics},
  volume = {310},
  number = {2},
  pages = {493--551},
  doi = {10.1016/j.aop.2004.01.004},
  langid = {english}
}

@article{bergmanBandTouchingRealspace2008,
  title = {Band Touching from Real-Space Topology in Frustrated Hopping Models},
  author = {Bergman, Doron L. and Wu, Congjun and Balents, Leon},
  year = 2008,
  month = sep,
  journal = {Physical Review B},
  volume = {78},
  number = {12},
  pages = {125104},
  doi = {10.1103/PhysRevB.78.125104},
  langid = {english}
}

@article{bondersonQuasiTopologicalPhasesMatter2013,
  title = {Quasi-{{Topological Phases}} of {{Matter}} and {{Topological Protection}}},
  author = {Bonderson, Parsa and Nayak, Chetan},
  year = 2013,
  month = may,
  journal = {Physical Review B},
  volume = {87},
  number = {19},
  eprint = {1212.6395},
  pages = {195451},
  doi = {10.1103/PhysRevB.87.195451},
  archiveprefix = {arXiv}
}

@article{castelnovoQuantumMechanicsClassical2005,
  title = {From Quantum Mechanics to Classical Statistical Physics: {{Generalized Rokhsar}}--{{Kivelson Hamiltonians}} and the ``{{Stochastic Matrix Form}}'' Decomposition},
  shorttitle = {From Quantum Mechanics to Classical Statistical Physics},
  author = {Castelnovo, Claudio and Chamon, Claudio and Mudry, Christopher and Pujol, Pierre},
  year = 2005,
  month = aug,
  journal = {Annals of Physics},
  volume = {318},
  number = {2},
  pages = {316--344},
  doi = {10.1016/j.aop.2005.01.006},
  langid = {english}
}

@article{changEntanglementSpectrumEntropy2020,
  title = {Entanglement Spectrum and Entropy in Topological Non-{{Hermitian}} Systems and Non-Unitary Conformal Field Theories},
  author = {Chang, Po-Yao and You, Jhih-Shih and Wen, Xueda and Ryu, Shinsei},
  year = 2020,
  month = jul,
  journal = {Physical Review Research},
  volume = {2},
  number = {3},
  eprint = {1909.01346},
  pages = {033069},
  doi = {10.1103/PhysRevResearch.2.033069},
  archiveprefix = {arXiv}
}

@article{chenScalingEntanglement2+1dimensional2015,
  title = {Scaling of Entanglement in 2+1-Dimensional Scale-Invariant Field Theories},
  author = {Chen, Xiao and Cho, Gil Young and Faulkner, Thomas and Fradkin, Eduardo},
  year = 2015,
  month = feb,
  journal = {Journal of Statistical Mechanics: Theory and Experiment},
  volume = {2015},
  number = {2},
  eprint = {1412.3546},
  pages = {P02010},
  doi = {10.1088/1742-5468/2015/02/P02010},
  archiveprefix = {arXiv},
  langid = {american}
}

@article{creutzigLogarithmicConformalField2013,
  title = {Logarithmic {{Conformal Field Theory}}: {{Beyond}} an {{Introduction}}},
  shorttitle = {Logarithmic {{Conformal Field Theory}}},
  author = {Creutzig, Thomas and Ridout, David},
  year = 2013,
  month = dec,
  journal = {Journal of Physics A: Mathematical and Theoretical},
  volume = {46},
  number = {49},
  eprint = {1303.0847},
  pages = {494006},
  doi = {10.1088/1751-8113/46/49/494006},
  archiveprefix = {arXiv}
}

@article{dharSelforganizedCriticalState1990,
  title = {Self-Organized Critical State of Sandpile Automaton Models},
  author = {Dhar, Deepak},
  year = 1990,
  month = apr,
  journal = {Physical Review Letters},
  volume = {64},
  number = {14},
  pages = {1613--1616},
  doi = {10.1103/PhysRevLett.64.1613},
  langid = {english}
}

@book{difrancescoConformalFieldTheory1997,
  title = {Conformal {{Field Theory}}},
  author = {Di Francesco, Philippe and Mathieu, Pierre and S{\'e}n{\'e}chal, David},
  year = 1997,
  series = {Graduate {{Texts}} in {{Contemporary Physics}}},
  publisher = {Springer New York},
  doi = {10.1007/978-1-4612-2256-9},
  isbn = {978-1-4612-7475-9 978-1-4612-2256-9},
  langid = {english}
}

@article{dijkgraafRelatingFieldTheories2010,
  title = {Relating {{Field Theories}} via {{Stochastic Quantization}}},
  author = {Dijkgraaf, Robbert and Orlando, Domenico and Reffert, Susanne},
  year = 2010,
  month = jan,
  journal = {Nuclear Physics B},
  volume = {824},
  number = {3},
  eprint = {0903.0732},
  pages = {365--386},
  doi = {10.1016/j.nuclphysb.2009.07.018},
  archiveprefix = {arXiv}
}

@article{flohrBitsPiecesLogarithmic2003,
  title = {Bits and {{Pieces}} in {{Logarithmic Conformal Field Theory}}},
  author = {Flohr, Michael},
  year = 2003,
  month = oct,
  journal = {International Journal of Modern Physics A},
  volume = {18},
  number = {25},
  eprint = {hep-th/0111228},
  pages = {4497--4591},
  doi = {10.1142/S0217751X03016859},
  archiveprefix = {arXiv},
  langid = {american}
}

@book{fradkinFieldTheoriesCondensed2013,
  title = {Field {{Theories}} of {{Condensed Matter Physics}}},
  author = {Fradkin, Eduardo},
  year = 2013,
  edition = {2},
  publisher = {Cambridge University Press},
  doi = {10.1017/CBO9781139015509},
  isbn = {978-0-521-76444-5},
  langid = {american}
}

@misc{ginspargAppliedConformalField1988,
  title = {Applied {{Conformal Field Theory}}},
  author = {Ginsparg, Paul},
  year = 1988,
  month = nov,
  number = {arXiv:hep-th/9108028},
  eprint = {hep-th/9108028},
  publisher = {arXiv},
  doi = {10.48550/arXiv.hep-th/9108028},
  archiveprefix = {arXiv}
}

@article{gossetCorrelationLengthGap2016,
  title = {Correlation {{Length}} versus {{Gap}} in {{Frustration-Free Systems}}},
  author = {Gosset, David and Huang, Yichen},
  year = 2016,
  month = mar,
  journal = {Physical Review Letters},
  number = {9},
  pages = {097202},
  doi = {10.1103/PhysRevLett.116.097202},
  langid = {english}
}

@article{gossetLocalGapThreshold2016,
  title = {Local Gap Threshold for Frustration-Free Spin Systems},
  author = {Gosset, David and Mozgunov, Evgeny},
  year = 2016,
  month = sep,
  journal = {Journal of Mathematical Physics},
  volume = {57},
  number = {9},
  pages = {091901},
  doi = {10.1063/1.4962337}
}

@article{gurarieHaldaneRezayiQuantumHall1997,
  title = {The {{Haldane-Rezayi Quantum Hall State}} and {{Conformal Field Theory}}},
  author = {Gurarie, V. and Flohr, M. and Nayak, C.},
  year = 1997,
  month = aug,
  journal = {Nuclear Physics B},
  volume = {498},
  number = {3},
  eprint = {cond-mat/9701212},
  pages = {513--538},
  doi = {10.1016/S0550-3213(97)00351-9},
  archiveprefix = {arXiv},
  langid = {american}
}

@article{gurarieLogarithmicOperatorsConformal1993,
  title = {Logarithmic {{Operators}} in {{Conformal Field Theory}}},
  author = {Gurarie, V.},
  year = 1993,
  month = dec,
  journal = {Nuclear Physics B},
  volume = {410},
  number = {3},
  eprint = {hep-th/9303160},
  pages = {535--549},
  doi = {10.1016/0550-3213(93)90528-W},
  archiveprefix = {arXiv}
}

@article{haldaneSpinsingletWaveFunction1988,
  title = {Spin-Singlet Wave Function for the Half-Integral Quantum {{Hall}} Effect},
  author = {Haldane, F. D. M. and Rezayi, E. H.},
  year = 1988,
  month = mar,
  journal = {Physical Review Letters},
  volume = {60},
  number = {10},
  pages = {956--959},
  doi = {10.1103/PhysRevLett.60.956},
  langid = {english}
}

@article{henleyClassicalQuantumDynamics2004,
  title = {From Classical to Quantum Dynamics at {{Rokhsar}}--{{Kivelson}} Points},
  author = {Henley, C L},
  year = 2004,
  month = mar,
  journal = {Journal of Physics: Condensed Matter},
  volume = {16},
  number = {11},
  pages = {S891-S898},
  doi = {10.1088/0953-8984/16/11/045},
  langid = {english}
}

@article{isakovDynamicsConformalQuantum2011,
  title = {Dynamics at and near Conformal Quantum Critical Points},
  author = {Isakov, S. V. and Fendley, P. and Ludwig, A. W. W. and Trebst, S. and Troyer, M.},
  year = 2011,
  month = mar,
  journal = {Physical Review B},
  volume = {83},
  number = {12},
  eprint = {1012.3806},
  pages = {125114},
  doi = {10.1103/PhysRevB.83.125114},
  archiveprefix = {arXiv}
}

@article{ivashkevichCorrelationFunctionsDense1999,
  title = {Correlation {{Functions}} of {{Dense Polymers}} and C=-2 {{Conformal Field Theory}}},
  author = {Ivashkevich, E. V.},
  year = 1999,
  month = mar,
  journal = {Journal of Physics A: Mathematical and General},
  volume = {32},
  number = {9},
  eprint = {cond-mat/9801183},
  pages = {1691--1699},
  doi = {10.1088/0305-4470/32/9/015},
  archiveprefix = {arXiv}
}

@article{izmailianLogarithmicConformalField2005,
  title = {Logarithmic {{Conformal Field Theory}} and {{Boundary Effects}} in the {{Dimer Model}}},
  author = {Izmailian, N. {\relax Sh}. and Priezzhev, V. B. and Ruelle, Philippe and Hu, Chin-Kun},
  year = 2005,
  month = dec,
  journal = {Physical Review Letters},
  volume = {95},
  number = {26},
  pages = {260602},
  doi = {10.1103/PhysRevLett.95.260602},
  langid = {english}
}

@misc{kauschCuriositiesC21995,
  title = {Curiosities at C=-2},
  author = {Kausch, Horst G.},
  year = 1995,
  month = oct,
  number = {arXiv:hep-th/9510149},
  eprint = {hep-th/9510149},
  publisher = {arXiv},
  doi = {10.48550/arXiv.hep-th/9510149},
  archiveprefix = {arXiv},
  langid = {american}
}

@article{kauschSymplecticFermions2000,
  title = {Symplectic {{Fermions}}},
  author = {Kausch, Horst G.},
  year = 2000,
  month = sep,
  journal = {Nuclear Physics B},
  volume = {583},
  number = {3},
  eprint = {hep-th/0003029},
  pages = {513--541},
  doi = {10.1016/S0550-3213(00)00295-9},
  archiveprefix = {arXiv},
  langid = {american}
}

@article{kitaevAnyonsExactlySolved2006,
  title = {Anyons in an Exactly Solved Model and Beyond},
  author = {Kitaev, Alexei},
  year = 2006,
  month = jan,
  journal = {Annals of Physics},
  volume = {321},
  number = {1},
  eprint = {cond-mat/0506438},
  pages = {2--111},
  doi = {10.1016/j.aop.2005.10.005},
  archiveprefix = {arXiv},
  langid = {american}
}

@article{kitaevFaulttolerantQuantumComputation2003,
  title = {Fault-Tolerant Quantum Computation by Anyons},
  author = {Kitaev, A. Yu},
  year = 2003,
  month = jan,
  journal = {Annals of Physics},
  volume = {303},
  number = {1},
  eprint = {quant-ph/9707021},
  pages = {2--30},
  doi = {10.1016/S0003-4916(02)00018-0},
  archiveprefix = {arXiv}
}

@article{koshinoTransportBilayerGraphene2006,
  title = {Transport in {{Bilayer Graphene}}: {{Calculations}} within a Self-Consistent {{Born}} Approximation},
  shorttitle = {Transport in {{Bilayer Graphene}}},
  author = {Koshino, Mikito and Ando, Tsuneya},
  year = 2006,
  month = jun,
  journal = {Physical Review B},
  volume = {73},
  number = {24},
  eprint = {cond-mat/0606166},
  pages = {245403},
  doi = {10.1103/PhysRevB.73.245403},
  archiveprefix = {arXiv}
}

@article{levinStringnetCondensationPhysical2005,
  title = {String-Net Condensation: {{A}} Physical Mechanism for Topological Phases},
  shorttitle = {String-Net Condensation},
  author = {Levin, Michael A. and Wen, Xiao-Gang},
  year = 2005,
  month = jan,
  journal = {Physical Review B},
  volume = {71},
  number = {4},
  pages = {045110},
  publisher = {American Physical Society},
  doi = {10.1103/PhysRevB.71.045110}
}

@article{liPhysicsHigherOrbital2016,
  title = {Physics of Higher Orbital Bands in Optical Lattices: A Review},
  shorttitle = {Physics of Higher Orbital Bands in Optical Lattices},
  author = {Li, Xiaopeng and Liu, W Vincent},
  year = 2016,
  month = sep,
  journal = {Reports on Progress in Physics},
  volume = {79},
  number = {11},
  pages = {116401},
  publisher = {IOP Publishing},
  doi = {10.1088/0034-4885/79/11/116401},
  langid = {english}
}

@article{liuSpontaneousSymmetryBreaking2010,
  title = {Spontaneous Symmetry Breaking in a Two-Dimensional Kagome Lattice},
  author = {Liu, Qin and Yao, Hong and Ma, Tianxing},
  year = 2010,
  month = jul,
  journal = {Physical Review B},
  volume = {82},
  number = {4},
  pages = {045102},
  doi = {10.1103/PhysRevB.82.045102},
  langid = {english}
}

@misc{masaokaFrustrationfreeFreeFermions2025,
  title = {Frustration-Free Free Fermions and Beyond},
  author = {Masaoka, Rintaro and Ono, Seishiro and Po, Hoi Chun and Watanabe, Haruki},
  year = 2025,
  month = mar,
  number = {arXiv:2503.12879},
  eprint = {2503.12879},
  publisher = {arXiv},
  doi = {10.48550/arXiv.2503.12879},
  archiveprefix = {arXiv}
}

@misc{masaokaRigorousLowerBound2024,
  title = {Rigorous Lower Bound of Dynamic Critical Exponents in Critical Frustration-Free Systems},
  author = {Masaoka, Rintaro and Soejima, Tomohiro and Watanabe, Haruki},
  year = 2024,
  month = jun,
  number = {arXiv:2406.06415},
  eprint = {2406.06415},
  publisher = {arXiv},
  doi = {10.48550/arXiv.2406.06415},
  archiveprefix = {arXiv}
}

@article{mccannLandauLevelDegeneracy2006,
  title = {Landau Level Degeneracy and Quantum {{Hall}} Effect in a Graphite Bilayer},
  author = {McCann, Edward and Fal'ko, Vladimir I.},
  year = 2006,
  month = mar,
  journal = {Physical Review Letters},
  volume = {96},
  number = {8},
  eprint = {cond-mat/0510237},
  pages = {086805},
  doi = {10.1103/PhysRevLett.96.086805},
  archiveprefix = {arXiv}
}

@article{murrayRenormalizationGroupStudy2014,
  title = {Renormalization Group Study of Interaction-Driven Quantum Anomalous {{Hall}} and Quantum Spin {{Hall}} Phases in Quadratic Band Crossing Systems},
  author = {Murray, James M. and Vafek, Oskar},
  year = 2014,
  month = may,
  journal = {Physical Review B},
  volume = {89},
  number = {20},
  pages = {201110},
  publisher = {American Physical Society},
  doi = {10.1103/PhysRevB.89.201110}
}

@article{nayakNonAbelianAnyonsTopological2008,
  title = {Non-{{Abelian}} Anyons and Topological Quantum Computation},
  author = {Nayak, Chetan and Simon, Steven H. and Stern, Ady and Freedman, Michael and Das Sarma, Sankar},
  year = 2008,
  month = sep,
  journal = {Reviews of Modern Physics},
  volume = {80},
  number = {3},
  pages = {1083--1159},
  doi = {10.1103/RevModPhys.80.1083},
  langid = {english}
}

@article{ngStateOperatorCorrespondence2013,
  title = {State/Operator Correspondence in Higher-Spin {{dS}}/{{CFT}}},
  author = {Ng, Gim Seng and Strominger, Andrew},
  year = 2013,
  month = may,
  journal = {Classical and Quantum Gravity},
  volume = {30},
  number = {10},
  pages = {104002},
  publisher = {IOP Publishing},
  doi = {10.1088/0264-9381/30/10/104002},
  langid = {english}
}

@article{olschlagerTopologicallyInducedAvoided2012,
  title = {Topologically {{Induced Avoided Band Crossing}} in an {{Optical Checkerboard Lattice}}},
  author = {{\"O}lschl{\"a}ger, Matthias and Wirth, Georg and Kock, Thorge and Hemmerich, Andreas},
  year = 2012,
  month = feb,
  journal = {Physical Review Letters},
  volume = {108},
  number = {7},
  pages = {075302},
  doi = {10.1103/PhysRevLett.108.075302},
  langid = {english}
}

@misc{onoFrustrationfreeFreeFermions2025,
  title = {Frustration-Free Free Fermions},
  author = {Ono, Seishiro and Masaoka, Rintaro and Watanabe, Haruki and Po, Hoi Chun},
  year = 2025,
  month = mar,
  number = {arXiv:2503.14312},
  eprint = {2503.14312},
  publisher = {arXiv},
  doi = {10.48550/arXiv.2503.14312},
  archiveprefix = {arXiv}
}

@article{parisiPERTURBATIONTHEORYGAUGE1981,
  title = {{{PERTURBATION THEORY WITHOUT GAUGE FIXING}}},
  author = {Parisi, G. and Wu, Yongshi},
  year = 1981,
  month = apr,
  journal = {Scientia Sinica},
  volume = {24},
  number = {4},
  pages = {483},
  langid = {american}
}

@article{pujariInteractionInducedDiracFermions2016,
  title = {Interaction-{{Induced Dirac Fermions}} from {{Quadratic Band Touching}} in {{Bilayer Graphene}}},
  author = {Pujari, Sumiran and Lang, Thomas C. and Murthy, Ganpathy and Kaul, Ribhu K.},
  year = 2016,
  month = aug,
  journal = {Physical Review Letters},
  volume = {117},
  number = {8},
  pages = {086404},
  publisher = {American Physical Society},
  doi = {10.1103/PhysRevLett.117.086404}
}

@article{rhimClassificationFlatBands2019,
  title = {Classification of Flat Bands According to the Band-Crossing Singularity of {{Bloch}} Wave Functions},
  author = {Rhim, Jun-Won and Yang, Bohm-Jung},
  year = 2019,
  month = jan,
  journal = {Physical Review B},
  volume = {99},
  number = {4},
  pages = {045107},
  publisher = {American Physical Society},
  doi = {10.1103/PhysRevB.99.045107}
}

@article{rhimSingularFlatBands2021,
  title = {Singular Flat Bands},
  author = {Rhim, Jun-Won and {and Yang}, Bohm-Jung},
  year = 2021,
  month = jan,
  journal = {Advances in Physics: X},
  volume = {6},
  number = {1},
  pages = {1901606},
  publisher = {Taylor \& Francis},
  doi = {10.1080/23746149.2021.1901606},
  langid = {american}
}

@article{rokhsarSuperconductivityQuantumHardCore1988,
  title = {Superconductivity and the {{Quantum Hard-Core Dimer Gas}}},
  author = {Rokhsar, Daniel S. and Kivelson, Steven A.},
  year = 1988,
  month = nov,
  journal = {Physical Review Letters},
  volume = {61},
  number = {20},
  pages = {2376--2379},
  publisher = {American Physical Society},
  doi = {10.1103/PhysRevLett.61.2376},
  langid = {american}
}

@article{ruelleLogarithmicConformalInvariance2013,
  title = {Logarithmic Conformal Invariance in the {{Abelian}} Sandpile Model},
  author = {Ruelle, Philippe},
  year = 2013,
  month = jan,
  journal = {Journal of Physics A: Mathematical and Theoretical},
  volume = {46},
  number = {49},
  pages = {494014},
  publisher = {IOP Publishing},
  doi = {10.1088/1751-8113/46/49/494014},
  langid = {english}
}

@article{saleurPolymersPercolationTwo1992,
  title = {Polymers and Percolation in Two Dimensions and Twisted {{N}}=2 Supersymmetry},
  author = {Saleur, Hubert},
  year = 1992,
  month = sep,
  journal = {Nuclear Physics B},
  volume = {382},
  number = {3},
  eprint = {hep-th/9111007},
  pages = {486--531},
  doi = {10.1016/0550-3213(92)90657-W},
  archiveprefix = {arXiv}
}

@article{sengokuQuasilocalFrustrationFreeFree2025,
  title = {Quasi-Local {{Frustration-Free Free Fermions}}},
  author = {Sengoku, Shunsuke and Po, Hoi Chun and Watanabe, Haruki},
  year = 2025,
  month = sep,
  journal = {Physical Review B},
  volume = {112},
  number = {11},
  eprint = {2505.01010},
  pages = {115104},
  doi = {10.1103/794x-jdn7},
  archiveprefix = {arXiv}
}

@book{simonTopologicalQuantum2023,
  title = {Topological Quantum},
  author = {Simon, Steven H.},
  year = 2023,
  publisher = {Oxford university press},
  isbn = {978-0-19-888672-3},
  langid = {american},
  lccn = {006.384 3}
}

@article{sunTimereversalSymmetryBreaking2008,
  title = {Time-Reversal Symmetry Breaking and Spontaneous Anomalous {{Hall}} Effect in {{Fermi}} Fluids},
  author = {Sun, Kai},
  year = 2008,
  journal = {Physical Review B},
  volume = {78},
  number = {24},
  doi = {10.1103/PhysRevB.78.245122}
}

@article{sunTopologicalInsulatorsNematic2009,
  title = {Topological {{Insulators}} and {{Nematic Phases}} from {{Spontaneous Symmetry Breaking}} in {{2D Fermi Systems}} with a {{Quadratic Band Crossing}}},
  author = {Sun, Kai and Yao, Hong and Fradkin, Eduardo and Kivelson, Steven A.},
  year = 2009,
  month = jul,
  journal = {Physical Review Letters},
  volume = {103},
  number = {4},
  pages = {046811},
  doi = {10.1103/PhysRevLett.103.046811},
  langid = {english}
}

@article{sunTopologicalSemimetalFermionic2012,
  title = {Topological Semimetal in a Fermionic Optical Lattice},
  author = {Sun, Kai and Liu, W. Vincent and Hemmerich, Andreas and Das Sarma, S.},
  year = 2012,
  month = jan,
  journal = {Nature Physics},
  volume = {8},
  number = {1},
  pages = {67--70},
  publisher = {Nature Publishing Group},
  doi = {10.1038/nphys2134},
  langid = {english}
}

@article{uebelackerInstabilitiesQuadraticBand2011,
  title = {Instabilities of Quadratic Band Crossing Points},
  author = {Uebelacker, Stefan and Honerkamp, Carsten},
  year = 2011,
  month = nov,
  journal = {Physical Review B},
  volume = {84},
  number = {20},
  pages = {205122},
  doi = {10.1103/PhysRevB.84.205122},
  langid = {english}
}

@article{verstraeteCriticalityAreaLaw2006,
  title = {Criticality, the {{Area Law}}, and the {{Computational Power}} of {{Projected Entangled Pair States}}},
  author = {Verstraete, F. and Wolf, M. M. and {Perez-Garcia}, D. and Cirac, J. I.},
  year = 2006,
  month = jun,
  journal = {Physical Review Letters},
  volume = {96},
  number = {22},
  pages = {220601},
  doi = {10.1103/PhysRevLett.96.220601},
  langid = {english}
}

@article{zengTuningTopologicalPhase2018,
  title = {Tuning Topological Phase and Quantum Anomalous {{Hall}} Effect by Interaction in Quadratic Band Touching Systems},
  author = {Zeng, Tian-Sheng and Zhu, Wei and Sheng, Donna},
  year = 2018,
  month = sep,
  journal = {npj Quantum Materials},
  volume = {3},
  number = {1},
  pages = {49},
  publisher = {Nature Publishing Group},
  doi = {10.1038/s41535-018-0120-5},
  langid = {english}
}

@article{zhuInteractionDrivenSpontaneousQuantum2016,
  title = {Interaction-{{Driven Spontaneous Quantum Hall Effect}} on a {{Kagome Lattice}}},
  author = {Zhu, W. and Gong, Shou-Shu and Zeng, Tian-Sheng and Fu, Liang and Sheng, D. N.},
  year = 2016,
  month = aug,
  journal = {Physical Review Letters},
  volume = {117},
  number = {9},
  pages = {096402},
  doi = {10.1103/PhysRevLett.117.096402},
  langid = {english}
}

\end{document}